\documentclass[12pt, a4paper]{article}

\usepackage[height=25cm,width=16cm]{geometry}
\usepackage{amsfonts}
\usepackage{amsmath}
\usepackage{amssymb}
\usepackage{ascmac}
\usepackage{dcolumn}
\usepackage{bm,here}
\usepackage{subfig}
\usepackage{comment}
\usepackage{ifpdf}
\usepackage{slashed}
\usepackage{colortbl}
\usepackage{color}
\usepackage[mathscr]{eucal}
\usepackage[charter]{mathdesign}

\ifpdf        
  \usepackage{graphicx}     
  \usepackage[bookmarksopen,colorlinks=true,linkcolor=bblue,citecolor=ppink,urlcolor=ppink,linktocpage=false]{hyperref}
\else     
  \usepackage[dvipdfmx]{graphicx}     
  \usepackage[dvipdfmx,bookmarksopen,colorlinks=true,linkcolor=bblue,citecolor=ppink,urlcolor=ppink]{hyperref}
\fi

\usepackage{multicol}
\definecolor{red}{rgb}{1,0,0}
\definecolor{ppink}{rgb}{0.921545,0.440586,0.687243}
\definecolor{bblue}{rgb}{0.400000,0.400000,1.000000}

\makeatletter
\@addtoreset{equation}{section}

\makeatother

\begin{document}


\newcommand{\vev}[1]{ \left\langle {#1} \right\rangle }
\newcommand{\bra}[1]{ \langle {#1} | }
\newcommand{\ket}[1]{ | {#1} \rangle }
\newcommand{\EV}{ \ {\rm eV} }
\newcommand{\KEV}{ \ {\rm keV} }
\newcommand{\MEV}{\  {\rm MeV} }
\newcommand{\GEV}{\  {\rm GeV} }
\newcommand{\TEV}{\  {\rm TeV} }
\newcommand{\1}{\mbox{1}\hspace{-0.25em}\mbox{l}}
\newcommand{\Red}[1]{{\color{red} {#1}}}

\newcommand{\prn}[1]{\left( {#1} \right)}
\newcommand{\com}[1]{\left[ {#1} \right]}
\newcommand{\lmk}{\left(}  
\newcommand{\rmk}{\right)}
\newcommand{\lkk}{\left[}  
\newcommand{\rkk}{\right]}
\newcommand{\lhk}{\left \{ }  
\newcommand{\rhk}{\right \} }
\newcommand{\del}{\partial}  
\newcommand{\la}{\left\langle} 
\newcommand{\ra}{\right\rangle}
\newcommand{\half}{\frac{1}{2}}

\newcommand{\dd}{\mathrm{d}}
\newcommand{\Mpl}{M_{\rm Pl}}
\newcommand{\mg}{m_{3/2}}
\newcommand{\abs}[1]{\left\vert {#1} \right\vert}
\newcommand{\mphi}{m_I}
\newcommand{\Hz}{\ {\rm Hz}}
\newcommand{\Min}{{\rm Min}}
\newcommand{\Max}{{\rm Max}}
\newcommand{\Kahler}{K\"{a}hler }
\newcommand{\cphi}{\varphi}

\newcommand{\qel}{\hat{q}_{\rm el}}
\newcommand{\ksplit}{k_{\rm split}}
\def\GDM{\Gamma_{\rm DM}}
\def\Gsplit{\Gamma_{\rm split}}

\def\mg{m_{3/2}}
\def\Im{{\rm Im}}
\def\bea{\begin{array}}
\def\eea{\end{array}}
\newcommand{\beq}{\begin{eqnarray}}
\newcommand{\eeq}{\end{eqnarray}}
\def\Mpl{M_{\rm pl}}
\def\Td{T_{\rm decay}}
\def\Gphi{\tilde{\Gamma}_I}
\def\ti{\tilde{t}}
\def\sp{_{\rm split}}
\def\for{\qquad {\rm for}}


\begin{titlepage}

\begin{flushright}
IPMU 15-0096
\end{flushright}

\vskip 1.in
\begin{center}
{\huge \bf 
Thermalization Process after Inflation\\[.2em]
{\LARGE and\\[.7em]
Effective Potential of Scalar Field}
}
\vskip .65in

{\Large Kyohei Mukaida$^{\lozenge}$ 
and Masaki Yamada$^{\blacklozenge}$ 
}


\vskip .35in
\begin{tabular}{ll}
$^{\lozenge}$ &\!\! {\em Kavli IPMU (WPI), UTIAS,}\\
&{\em The University of Tokyo, Kashiwa, Chiba 277-8583, Japan}\\[.3em]
$^{\blacklozenge}$ &\!\! {\em Institute for Cosmic Ray Research, }\\
&{\em The University of Tokyo, Kashiwa, Chiba 277-8582, Japan}
\end{tabular}

\vskip .75in

\begin{abstract}  
\noindent
We investigate the thermalization process of the Universe after inflation 
to determine the evolution of the effective temperature. 
The time scale of thermalization is found to be so long that it delays the evolution of the effective temperature, 
and the resulting maximal temperature of the Universe can be significantly lower than the one obtained in the literature. 
Our results clarify 
the finite density corrections to the effective potential of a scalar field
and also processes of heavy particle production.
In particular, 
we find that the maximum temperature of the Universe may be at most 
electroweak scale if the reheating temperature is as low as ${\cal O} (1) \MEV$, which implies that
the electroweak symmetry may be marginally restored.
In addition, it is noticeable that the dark matter may not be produced from thermal plasma in such a low reheating scenario,
since the maximum temperature can be smaller than the conventional estimation by five orders of magnitude.
We also give implications to the Peccei-Quinn mechanism and the Affleck-Dine baryogenesis. 
\end{abstract}

\end{center}
\end{titlepage}

\tableofcontents
\newpage

\section{\label{sec1}Introduction and Summary}

\subsection{\label{sec1-1}Introduction}

Slow roll inflationary scenarios are 
successful 
in light of the solution to the horizon problem, flatness problem, 
and the origin of the large scale structure. 
Inflation is usually driven by a finite energy density of 
a slowly rolling scalar field, called inflaton. 
After the slow roll conditions fail, inflaton starts to oscillate around its potential minimum 
and the energy density of the Universe is then dominated by that of the oscillating inflaton. 
In order to proceed the big bang nucleosynthesis, 
the inflaton has to decay into radiation, 
which is referred to as reheating~\cite{Linde:1981mu, Albrecht:1982mp}.

While the inflaton elegantly solves the above problems 
and provides the primordial density perturbations, 
there still remains unsolved cosmological issues;
dark matter and baryon asymmetry of the Universe.
Scalar fields other than the inflaton may play essential roles in solving these remaining problems.
For instance, 
dark matter can be explained by a pseudo-NG boson, called axion~\cite{Weinberg:1977ma}, 
which is associated with the spontaneously symmetry breaking (SSB) 
of PQ symmetry triggered by a PQ-charged scalar field~\cite{Peccei:1977hh,Peccei:1977ur}. 
In supersymmetric (SUSY) theories, 
the Affleck-Dine mechanism can explain 
the origin of baryon asymmetry by using a baryonic scalar field~\cite{Affleck:1984fy, Dine:1995kz}. 
Generally, scalar fields should have interaction terms with radiation 
in order to successfully lose their energy, 
otherwise they tend to dominate the energy density of the Universe. 
This very interaction makes the dynamics of scalar fields non-trivial.

One of prominent effects  caused by the interaction between the scalar fields and radiation
is modification of their effective potential.\footnote{
	For other relevant effects of background thermal plasma on scalar condensates in the Universe,
	see Refs.~\cite{Yokoyama:2004pf,Mukaida:2012qn,Mukaida:2012bz,Mukaida:2013xxa,Drewes:2013iaa,Cheung:2015iqa} for instance.
}
As the thermal plasma grows after inflation, 
it drastically affects 
the dynamics of scalar fields.
At a sufficiently high temperature, for instance, 
thermal effects may induce a positive thermal mass for a scalar field 
and make it stay at the 
point of the potential where particles coupled to the scalar
field remain massless.
When the scalar field is responsible to the SSB 
of some symmetry, like the Higgs boson 
or a PQ breaking scalar field, 
the symmetry is restored at that high temperature~\cite{Kirzhnits:1972iw, Kirzhnits:1972ut, Dolan:1973qd, Weinberg:1974hy, Kirzhnits:1976ts}. 
Then, a phase transition occurs 
when the thermal mass decreases down to the zero-temperature mass of the boson field. 
At the phase transition, 
topological defects, such as cosmic strings and domain walls, may form 
and affect the evolution of the Universe. 
In QCD axion models, 
there may be the domain wall problem when the PQ symmetry is broken after inflation~\cite{Zeldovich:1974uw, Sikivie:1982qv}, 
while 
there is the isocurvature problem when it is broken before inflation~\cite{Axenides:1983hj, Seckel:1985tj, Turner:1990uz}. 
The abundance of axion DM also depends on whether the PQ symmetry is broken before or after inflation. 
In SUSY theories, 
the Affleck-Dine mechanism can 
generate baryon asymmetry by using a scalar field carrying a nonzero baryon charge~\cite{Affleck:1984fy, Dine:1995kz}. 
The amount of baryon asymmetry is sometimes affected by the finite temperature effects~\cite{Dine:1995kz, Allahverdi:2000zd, Asaka:2000nb, Fujii:2001zr,Anisimov:2000wx}. 
These examples show that 
the thermal effects on boson fields drastically affect the evolutions of their dynamics 
and change their predictions. 
Therefore, we should clarify the evolution of the temperature after inflation.

One might assume that the reheating occurs instantaneously 
and regard the reheating temperature as the maximum temperature of the Universe. 
More carefully, one may solve the Boltzmann equation 
for the inflaton and radiation 
to include the production of radiation before the complete decay of inflaton~\cite{Chung:1998rq, Giudice:2000ex}. 
Even in this method, 
``{\it instantaneous thermalization}'' of radiation is implicitly assumed. 
In Ref.~\cite{Harigaya:2013vwa}, however, we have pointed out that 
the time scale of thermalization of radiation is finite 
and affects the reheating process of the Universe after inflation.\footnote{
	See also Refs.~\cite{Davidson:2000er,Allahverdi:2002pu}.
}

In this paper, 
we further investigate the thermalization process of inflaton decay products 
and provides the evolution of an effective temperature which describes the strength of 
finite temperature effects.
We find that 
the maximal temperature of the Universe after inflation 
is much smaller than that obtained by the ``{\it instantaneous thermalization}'' assumption,
by extending the analysis given in Ref.~\cite{Harigaya:2013vwa}
to the era before the soft sector is thermalized.
The discrepancy becomes larger for smaller decay rate of inflaton 
and can be as large as about five order of magnitude.

Our results clarify the finite density corrections to the effective potential of the scalar field
such as so called the thermal mass and the thermal log potential;
which are essential so as to determine the beginning time of oscillation of the scalar field
and also the condition of the symmetry restoration by finite density corrections.
For example, 
we find that 
the maximum temperature of the Universe may be at most $\sim 100\text{GeV}$
if the reheating temperature is as low as ${\cal O} (1) \MEV$.
In this case, the electroweak phase transition is marginally restored after inflation.
We also find that the reheating temperature may have to be two order of magnitude lower than
the PQ breaking scale in order to avoid the restoration of PQ symmetry. 
In addition, we investigate the effect on the AD field, which generates baryon asymmetry 
by the Affleck-Dine mechanism. 
We find that 
the result is consistent with the case that one assumes ``{\it instantaneous thermalization}'' 
when the VEV of the AD field is larger than the mass of inflaton.

In order to study the finite density corrections to the effective potential,
we have to discuss ``heavy'' particle production processes with the mass $\sim |\phi|$.
Hence, our discussion can be also applied to heavy DM production processes.
Although in the literature it was expected that DM can be produced from the thermal plasma 
before reheating is completed~\cite{Chung:1998rq, Giudice:2000ex}, 
our result implies that they may not be produced in a low reheating temperature scenario. 
Instead, DM may be efficiently produced via scatterings between the soft thermalized plasma and 
hard primaries as studied in Ref.~\cite{Harigaya:2014waa},
or directly produced via inflaton decay.

In the next section, we briefly explain the finite density effects on boson fields,
and derive the relation between the effective temperature which characterizes the finite density corrections
to the effective potential
and the distribution function of radiation. 
Then we investigate the evolution of the thermalization process 
and calculate the effective temperature of radiation in Sec.~\ref{thermalization}.  
In Sec.~\ref{applications}, we apply our results to some important scenarios, 
such as the restoration of PQ symmetry and electroweak symmetry, 
and the Affleck-Dine baryogenesis. 
Section~\ref{conclusion} is devoted to the conclusion.

\subsection{\label{sec1-2}Summary of our results}

Before we explain the detail of our calculations, 
here we summarize our main results. 
If a system is not in thermal equilibrium,
we cannot characterize the system simply by a temperature in general.
In this paper, 
we define an effective temperature $T_*$ such that it describes the finite density effect on the potential of scalar fields
because our main purpose is to clarify the effects of radiation on scalar fields.
Note that these scalar fields of our interests are different from inflaton,
and that inflaton is assumed to reheat the Universe via Planck-suppressed decays throughout this paper.\footnote{
	If inflaton reheats the Universe via not so small couplings,
	the preheating and thermal dissipations after that may become important.
	See Refs.~\cite{Kofman:1994rk,Kofman:1997yn,Mukaida:2012bz} for instance.
}
See Sec.~\ref{sec:setup} for more details of our setup.

Let us first summarize important equations.
If $\chi_\phi$-particles, 
which are particles interacting with a scalar field $\phi$, are not directly produced from inflaton decay, 
the thermal potential of $\phi$ is roughly given by 
\begin{align}
	 V_{\rm eff} (\phi) \sim 
	\begin{cases}
		 \alpha_\phi T_*^2 \phi^2 
		 &\text{for}~~ 
		 | g_ \phi \phi | \ll T_* \\[.5em]
		 \alpha^2 T_*^4 \log \lmk \cfrac{\phi^2}{T_*^2} \rmk 
		 &\text{for}~~ 
		 |g_\phi \phi | \gg T_*
	\end{cases},
\end{align}
where $\alpha_\phi (\equiv g_\phi^2 / 4\pi)$ denotes a typical coupling between the scalar field
and $\chi_\phi$-fields [See Eq.~\eqref{eq:int_phi_rad} and Fig.~\ref{fig:schm}],
and $\alpha$ is the relevant coupling constant for thermalization process.

In Sec.~\ref{thermalization}, 
we show that the evolution of the effective temperature for the scalar field $\phi$
is dramatically different from that of conventional studies.
The evolution of the effective temperature $T_*$ after inflation is found to be 
\begin{align}
	\frac{T_*}{\mphi} \sim 
	\begin{cases}
		\alpha^{1/2} \Gphi^{1/2} \lmk \cfrac{\ti}{\ti_{\rm ini}} \rmk^{-1/4} &\text{for}~~  
		\ti_{\rm ini} \lesssim \ti \lesssim \ti_{\rm soft} \\[1em]
		\alpha \Gphi^{1/2} \lmk \cfrac{\ti}{\ti_{\rm soft}} \rmk 
		&\text{for}~~  \ti_{\rm soft} \lesssim \ti \lesssim \ti_{\rm max} \\[1em]
		\alpha^{4/5} \Gphi^{2/5} \lmk \cfrac{\ti}{\ti_{\rm max}} \rmk^{-1/4} 
		&\text{for}~~  \ti_{\rm max}  \lesssim \ti \lesssim \ti_{\rm RH} 
	\end{cases}
\end{align}
where $m_I$ is inflaton mass, $\alpha$ is the relevant coupling constant for thermalization process, 
$\Gphi$ is the decay rate of inflaton normalized by that of dimension-five operator, 
and $\ti$ is a cosmic time normalized by the inflaton mass. 
The time scales are given by 
\begin{align}
	 \ti_{\rm ini} &\equiv \alpha^{-1} \Gphi^{-1/2}, \\
	 \ti_{\rm soft} &\equiv \alpha^{-3} \Gphi^{-1/2}, \\
	 \ti_{\rm max} &\equiv \alpha^{-16/5} \Gphi^{-3/5}, \\
	 \ti_{\rm RH} &\equiv \Gphi^{-1} \frac{\Mpl^2}{m_I^2}, 
\end{align}
where $\Mpl$ ($\simeq 2.4 \times 10^{18} \GEV$) is the reduced Planck scale. 
We consider the case of $\Gphi \lesssim 1$ 
so that we can neglect non-perturbative effects of inflaton decay. 

To see a rough sketch,
we briefly explain the thermalization process after inflation. 
Figs.~\ref{distribution1}, \ref{distribution2} and \ref{distribution3} may be helpful
to understand the essence of the following discussion.
First, inflaton decays into hard primaries, whose energy is of order the mass of inflaton $m_I$. 
At the very first stage, a showering of hard primaries produces the soft population.
Soon after that, the hard primaries then inelastically scatter with each other 
and release their energy into soft particles via in-medium cascading processes. 
Due to the loss of causality, 
the soft thermal-like distribution with $k^{-1}$ 
emerges after the time defined by $\ti_{\rm ini} \equiv \alpha^{-1} \Gphi^{-1/2}$. 
The inelastic scattering rate is suppressed by the LPM effect, 
which is stronger in lower dense environment. 
Since the density of the hard primaries decreases with time due to the expansion of the Universe, 
the inelastic scattering rate decreases with time. 
Thus, the energy density of soft particles also decreases, 
so that their effective temperature $T_\ast$ decreases with time. 
Then, at $\ti \sim \ti_{\rm soft}$, 
soft particles are completely thermalized. 
At the same time, 
the number density of the Universe is dominated by the soft particles, 
so that the LPM suppression becomes weaker and weaker. 
Therefore the effective temperature of soft particles 
increases with time after $\ti \sim \ti_{\rm soft}$. 
Then, at $\ti \sim \ti_{\rm max}$, 
the primary particles can lose their energy completely and are thermalized within the Hubble time scale.
After that time, 
the temperature of the Universe can be defined definitely 
and is determined by the energy injected by the inflaton decay.%
\footnote{
Still, note that there remains a tail of cascading hard particles as can be seen from Fig.~\ref{distribution3}.
Though this tail does not contribute to neither the energy density nor the number density,
it can be a source of heavy particle production with masses of $m \gg T$~\cite{Harigaya:2014waa}.
}
Since the energy density of inflaton decreases with time, 
the temperature of the Universe decreases, too. 
Finally, at $\ti \sim \ti_{\rm RH}$, 
inflaton completely decays into radiation 
and reheating is completed.

There are two things to be noted. 
\begin{enumerate}
\item We cannot rely on the conventional estimation of the temperature, $T \sim \rho_r^{1/4}$, 
for $\tilde t \lesssim \tilde t_\text{max}$.
This is because the energy density is still dominated by the hard primaries for $\tilde t \lesssim \tilde t_\text{max}$,
whose distribution is far from thermal equilibrium.
\item The maximum temperature of the Universe may not be achieved before $\tilde t_\text{max}$,
and thus it may be significantly lower than the standard scenario in which the ``{\it instantaneous thermalization}'' is assumed. 
The effective temperature of the soft sector has two local maxima 
as $T_*/\mphi \sim \alpha^{1/2} \Gphi^{1/2}$ and $\alpha^{4/5} \Gphi^{2/5}$ 
at the time of 
$\ti = \ti_{\rm ini}$ and $\ti_{\rm max}$, respectively. 
In particular, for the case of $\Gphi \lesssim \alpha^3$, 
the effective temperature reaches its maximal value 
at $\ti \sim \ti_{\rm max}$ (see Fig.~\ref{fig2}): 
\begin{align}
		\alpha^{4/5} \Gphi^{2/5} \mphi
		~~~\text{for}~~ 
		\Gphi \lesssim \alpha^3. 
\end{align}
If the primary particles was thermalized instantaneously, 
the temperature of the plasma would be calculated from 
$T^4 \sim \rho_R \sim H \Gamma_\phi M_{\rm Pl}^2$~\cite{Chung:1998rq, Giudice:2000ex}. 
(This is shown as red dotted lines in Figs.~\ref{fig1} and \ref{fig2}, 
which clarifies that the thermal effect is overestimated 
in the ``{\it instantaneous thermalization}'' approximation.)
The most important quantity is the maximal temperature of the Universe 
because it determines whether a symmetry is restored after inflation or not. 
It is overestimated in the ``{\it instantaneous thermalization}'' approximation 
by the following factor: 
\begin{align}
	 \frac{T_{\rm max}^{\rm (inst)}}{\left. T_\ast \right|_{\rm max}^{\rm (this \ work)}} \sim 
	 \alpha^{-4/5} \Gphi^{-3/20} 
	 \lmk \frac{H_I}{m_I} \rmk^{1/4}, 
\end{align}
where we assume $\Gphi \lesssim \alpha^3$. 
This implies that the instantaneous thermalization approximation 
results in an overestimation
by about five order of magnitude 
for the case of $\alpha = 1/10$, $\Gphi = 10^{-27}$, and $H_I = m_I$, for example. 
Here note that the decay rate of inflaton is bounded from below 
not to spoil the success of BBN:
\begin{align}
	 \Gphi \sim \frac{T_{\rm RH}^2 \Mpl}{m_I^3} 
	 \sim 10^{-27} 
	 \lmk \frac{T_{\rm RH}}{1 \MEV} \rmk^2 
	 \lmk \frac{m_I}{10^{13} \GEV} \rmk^{-3}. 
\end{align}
\end{enumerate}

\section{\label{thermalization}Preliminaries}

\subsection{Hard primaries}
We focus on the era 
between the end of inflation and the completion of reheating, during which 
the energy density of the Universe is dominated by that of the inflaton oscillation. 
Under the quadratic potential of inflaton, 
its oscillation amplitude and energy density evolve as 
\begin{align}
	 I(t) &\simeq \lmk \frac{a (t_I)}{a(t)} \rmk^{3/2} e^{- \Gamma_I t/2} I_0  {\rm cos} (m_I t) \\
	 \rho_I &\simeq \lmk \frac{a (t_I)}{a(t)} \rmk^{3} e^{- \Gamma_I t} 3 H_I^2 \Mpl^2,  
\end{align}
respectively. 
Here, $a(t)$ is the scale factor, $t_I$ is the time at which inflation ends, 
$I_0$ is the initial amplitude of the inflaton oscillation, 
$\Gamma_I$ 
is inflaton decay rate, 
and $H_I$ is the Hubble parameter at the end of inflation. 
We define the following dimensionless parameters, $\Gphi$, $\ti$, as:
\begin{align}
	 \Gamma_I &\equiv \Gphi \frac{m_I^3}{\Mpl^2}, \\
	 \ti &\equiv m_I t. 
	 \label{f_h}
\end{align}
Note that we have $\ti \gtrsim 1$
because the inflaton has to oscillate at least once before it decays. 
We assume that the inflaton decays into radiation perturbatively. 
This assumption is fulfilled at least for $\Gphi \lesssim 1$. 
To see this, 
let us consider the case that the inflaton decays into a fermion $\chi_I$ through the interaction of $\lambda I \bar{\chi}_I \chi_I$. 
This interaction term gives an effective mass to the $\chi_I$ field, 
so that its frequency is given by $\omega_{\chi_I}^2 = k^2 + \lambda^2 I^2(t)$. 
The non-perturbative decay occurs when the adiabatic condition $|\dot{\omega}_{\chi_I} / \omega_{\chi_I}^2| \ll 1$ 
is violated. 
Since $\dot{I}|_{I \simeq 0} \lesssim m_I I_0$ and $k \simeq m_I/2$, 
the adiabatic condition can be rewritten as $\lambda I_0 \ll m_I$. 
Using the perturbative decay rate given by $\Gamma_I \sim \lambda^2 m_I$, 
we obtain the condition to the perturbative decay as $\Gphi \ll \Mpl^2 / I_0^2$. 
Since $I_0 \lesssim \Mpl$, 
the condition of perturbative decay is always satisfied at least for the case of $\Gphi \lesssim 1$. 
Otherwise, the inflaton might decay through non-perturbatively~\cite{Kofman:1994rk,Kofman:1997yn}, 
which is beyond the scope of this paper.

Since the energy of inflaton decay products 
(hereafter we call them as hard primaries) is 
of the order of inflaton mass $m_I$,
their phase space distribution is given as~\cite{McDonald:1999hd}
\begin{align}
	 f_h(t,p) \sim  \lmk p / \mphi \rmk^{-3/2} \Gphi \ti^{-1} \qquad {\rm for } \ \ti^{-2/3} \lesssim p / \mphi \lesssim 1, 
	 \label{hard-distribution}
\end{align}
where the factor in the parentheses represents the effect of redshift. 
It implies smaller number density but harder particles than those of thermal distribution
for relevant time scales in consideration; 
$f_h (t, m_I) \sim \tilde \Gamma_I \tilde t^{-1} \ll 1$. 
This is the reason why we call them hard primaries.
Obviously, number violating processes are essential in thermalization, and in fact
the hard primaries scatter with each other inelastically 
and emit soft particles. 
As we will see in the following, these soft particles take crucial roles
both in thermalization of hard primaries and in dynamics of a scalar field $\phi$
which is the main character in this paper.
Note that the species of soft particles can be different from that of hard primaries. 
For example, even if inflaton decays only into Standard-Model gauge bosons through 
Planck-suppressed operators, 
soft particles can be quarks and leptons. 
Thus, almost all of the fields interacting with a Standard Model particle are affected by the soft sector. 
In this sense, our calculation can be applied to wide range of models. 
How the soft particles are produced and affect the effective potential of $\phi$-condensate
is discussed in the subsequent sections.

\subsection{Setup}
\label{sec:setup}

\begin{figure}[t]
\centering 
\includegraphics[width=.45\textwidth
]{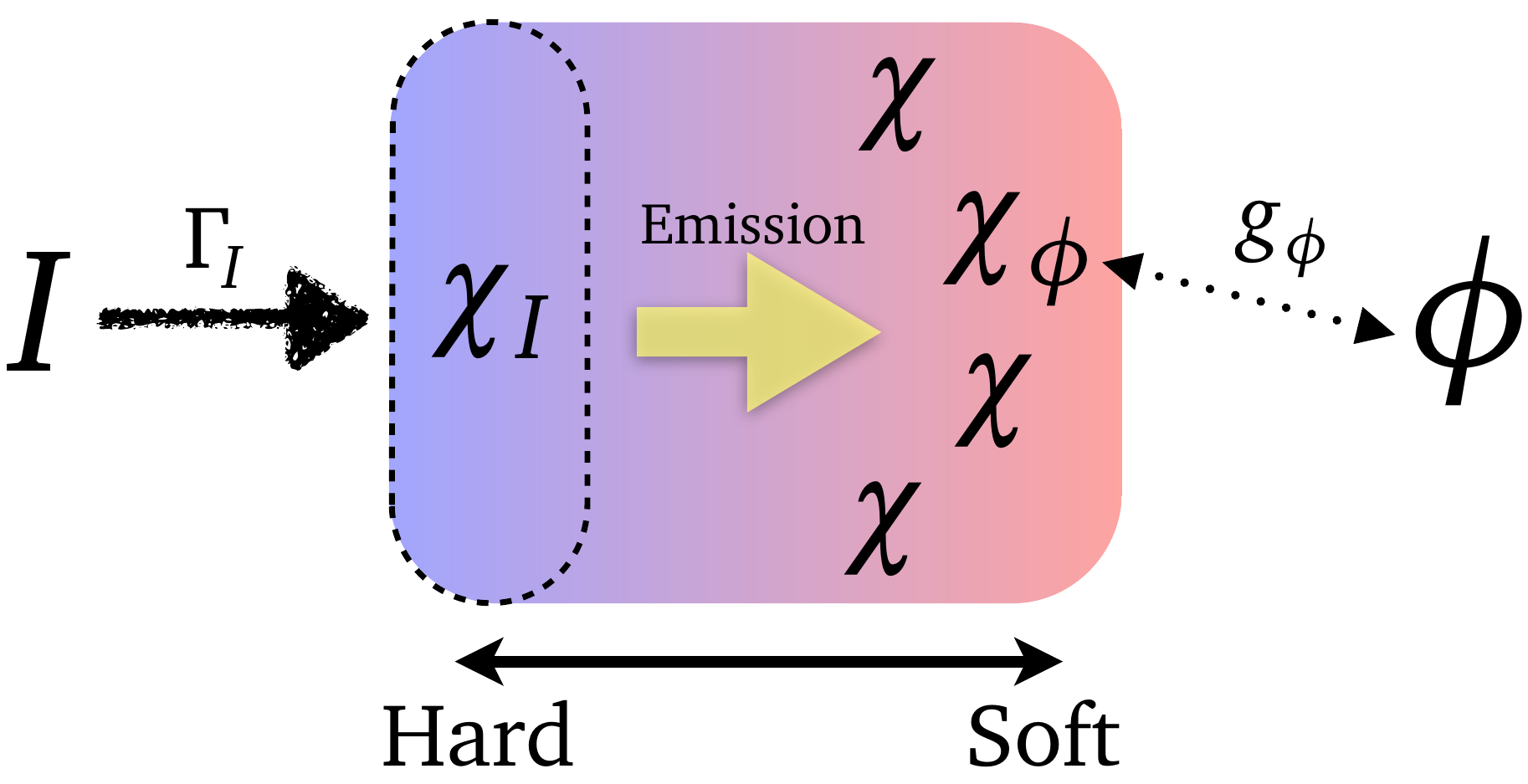} 
\caption{\small
{\bf Schematic figure of our setup}:
$I$ is the inflaton which creates hard primaries denoted by $\chi_I$,
which has typical momentum of $m_I$.
Hard primaries emit soft populations of SM particles $\chi$ (including $\chi_I$) 
and also $\chi_\phi$ (through SM interactions).
The field $\phi$ denotes the scalar field that we are interested in. It interacts with SM particles via $\chi_\phi$ 
with a typical coupling $g_\phi$.
}
  \label{fig:schm}
\end{figure}

Before going into details, let us clarify our setup.
Throughout this paper, we denote the scalar field in consideration 
(which is different from the inflaton) 
as $\phi$, 
and fields which couple to the scalar field and interact with standard model (SM) particles ({\it e.g.},~charged under SM gauge group)
as $\chi_\phi$.
In addition, inflaton, light fields directly produced via inflaton decay, and light fields other than $\chi_\phi$ are represented by
$I$, $\chi_I$, and $\chi$, respectively.
For simplicity, we assume that all the fields except for $\phi$ are charged under SM gauge group
with the predominant coupling being denoted as $g$,\footnote{
	The field $\phi$ itself can be either charged or singlet under SM gauge group, but
	note that if $\phi$ is charged under SM gauge group, then
	one should care that SM gauge group is not entirely broken down by the expectation value of $\phi$.
	Otherwise, the $t$-channel enhancement of massless gauge boson exchange is suppressed,
	and the following discussion should be altered.
}
and that masses of all the particles other than $\chi_\phi$ can be neglected.
That is, hard $\chi_I$-particles can emit themselves in general and hence $\chi$ includes $\chi_I$, 
but $\chi_\phi$ is treated separately since it can acquire a large effective mass of $|g_\phi \phi|$.
Here, symbolically, we represent the coupling between the scalar field $\phi$ and $\chi_\phi$
as $g_\phi$;
concrete examples are\footnote{
	For a complex scalar field theory, $\phi$ is regarded as its radial component.
}
\begin{align}
	-{\cal L}_\text{Yukawa} = g_\phi \phi \overline{\chi_{\phi \text{R}}} \chi_{\phi \text{L}} + \text{h.c.},~
	-{\cal L}_\text{quartic} = g_\phi^2 \phi^2 | \tilde \chi_\phi |^2,
	\label{eq:int_phi_rad}
\end{align}
for instance. 
Here, $\chi_{\phi i}$ and $\tilde \chi_\phi$ are fermion and boson, respectively, 
which are collectively denoted as $\chi_\phi$ hereafter. 
A schematic figure of our setup is shown in Fig.~\ref{fig:schm}.

At first, the scalar field condensates almost homogeneously with an expectation value~$\phi$.
Then, when the potential force overcomes the expansion of the Universe, 
$H \sim \sqrt{|\partial V_\text{eff}/\partial \phi^2|}$, 
it starts to roll down its effective potential $V_\text{eff}$ against the expansion of the Universe.
Eventually, it relaxes to  its ``equilibrium value'' at that time
by dissipating its energy into the background plasma.
Hence, it is of quite importance to understand the behavior of effective potential 
so as to know the cosmological fate of scalar condensates.
In particular, 
the effective potential is drastically changed 
via interactions with 
a dense and high temperature background plasma. 

\subsection{Thermal potentials}

\begin{figure}[t]
\centering 
\includegraphics[width=.45\textwidth
]{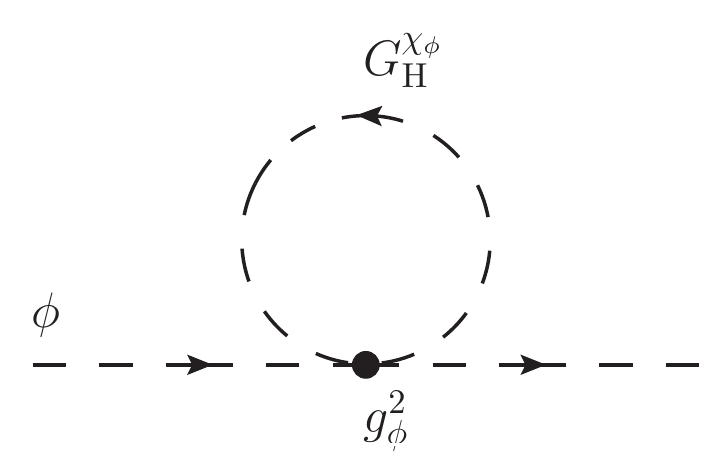} 
\caption{\small
Diagram responsible for the finite density correction to the mass term of $\phi$ 
in the case of quartic interaction. 
}
  \label{fig:diag1}
\end{figure}

There are two contributions to the effective potential depending on the expectation value
of $\phi$-condensation.
The first one comes from the abundant population of $\chi_\phi$-particles,
which typically has the following form:
\begin{align}
	V_\text{M} &\sim g_\phi^2 \phi^2 \int_{\bm p} \frac{f_{\chi_\phi} (\bm{p})}{E_{\bm p}}, 
	\label{eq:mass} \\
	\int_{\bm p} &\equiv \int \frac{\dd p^3}{(2\pi)^3}, 
\end{align}
where $f_{\chi_\phi}$ and $E_{\bm p}$ ($= \sqrt{ |\bm{p}|^2 + |g_\phi \phi|^2 }$) 
stand for the distribution function and the energy of $\chi_\phi$-particles, respectively.
For instance, in the case of quartic interaction, $g_\phi^2 \phi^2 \tilde \chi_\phi^2$,
the effective potential for the homogeneous condensate of $\phi$ encodes the following term 
(see Fig.~\ref{fig:diag1}):
\begin{align}
	g_\phi^2 \phi^2 \int_{P} G_\text{H}^{\chi_\phi} (P)
	\simeq g_\phi^2 \phi^2 \int_{P} \left[ 1 + 2 f_{\chi_\phi} (P) \right] G_\text{J}^{\chi_\phi} (P) 
	\simeq g_\phi^2 \phi^2 \int_{\bm p} \frac{1}{E_{\bm p}} \left[ 1 + 2 f_{\chi_\phi} (\bm{p}) \right],
	\label{eq:}
\end{align}
with $f_{\chi_\phi} (\bm{p}) \equiv f_{\chi_\phi} (E_{\bm p}, \bm{p})$.
Here $G_\text{H/J}^{\chi_\phi}$ is the Hadamard/Jordan propagator defined by the commutator/anti-commutator:
$G_\text{H/J}^{\chi_\phi} (x, y) \equiv \text{Tr} \left< [ \hat\chi_\phi (x), \hat\chi_\phi^\dag (y) ]_\pm \right>$.
In the first similarity, we have assumed the Kadanoff-Baym ansatz between the Hadamard and Jordan propagator,
and then assumed that the spectrum is dominated by particle-like excitations in the second similarity.
One can see that the finite density correction is reproduced aside from the quadratic divergence
that should be canceled by the counter term. A similar computation yields the finite density correction
from the Yukawa interaction, $ g_\phi \phi \overline{\chi_{\phi \text{R}}} \chi_{\phi \text{L}}$, which is essentially the same as Eq.~\eqref{eq:mass}.
The expression of Eq.~(\ref{eq:mass}) coincides with the well-known thermal mass,
if $f_{\chi_\phi}$ is given by the thermal distribution and if the effective mass of $\chi_\phi$ is smaller
than the temperature, $m_{\chi_\phi} \sim |g_\phi \phi| \ll T$.
Motivated by this observation, we define the effective temperature for $\chi_\phi$-particles as follows:
\begin{align}
	T_{\ast,\chi_\phi}^2 \equiv \int_{\bm p} \frac{f_{\chi_\phi} (\bm{p})}{E_{\bm p}}.
	\label{eq:eff_T_phi}
\end{align}
Although the distribution $f_{\chi_\phi}$ might be far from thermal equilibrium, 
we can estimate the effective potential for the scalar field $\phi$, 
which is caused by abundant $\chi_\phi$-particles. 
Imitating the expression at thermal equilibrium, we write down this effective mass contribution as
\begin{align}
	V_\text{M} \sim \alpha_\phi T_{\ast, \chi_\phi}^2 \phi^2,
	\label{eq:mass_2}
\end{align}
where $\alpha_\phi = g_\phi^2 / 4 \pi$. 

\begin{figure}[t]
\centering 
\includegraphics[width=.45\textwidth
]{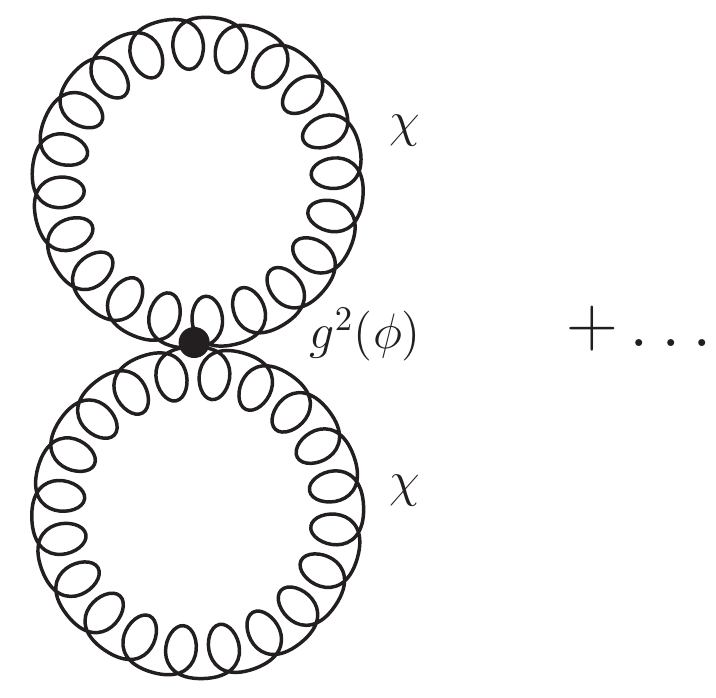} 
\caption{\small
Diagram responsible for the logarithmic correction to the effective potential of $\phi$. 
}
  \label{fig:diag2}
\end{figure}

When the field value of $\phi$-condensate is larger than the typical production scale of $\chi_\phi$-particles,
they are less likely to be produced from other light particles.
Hence, the contribution given in Eq.~\eqref{eq:mass} may be suppressed.
Even in this case, the effective potential for the $\phi$-condensate receives finite density corrections from
other light particles. 
This second contribution can be understood from the fact:
the runnings of coupling constants are affected by the expectation value of the scalar field. 
Suppose that $\phi$ has a large expectation value.
Since $\chi_\phi$ has a large effective mass of $|g_\phi \phi|$ and is less likely to be produced,
one can integrate out $\chi_\phi$-fields.
As a result, the running coupling constant of gauge group under which $\chi_\phi$ is charged
encodes the $\phi$-dependence as a threshold correction
of $\chi_\phi$-fields, or in other words, the $\phi$-condensate interacts with other light particles via
the operator $\sim g^{-2} (\phi)F^a_{\mu\nu}F^{a \mu \nu}$.\footnote{
	Other interactions like Yukawa may also encode the threshold correction of $\chi_\phi$.
}
Here we have assumed that $\phi$ is singlet under this gauge group.
This implies the following contribution to the effective potential for the $\phi$-condensate 
(see Fig.~\ref{fig:diag2}):\footnote{
Here we have assumed that $\chi_\phi$ is also charged under the dominant gauge group of coupling $g$.
If this is not the case, a factor should be multiplied,
which is defined as the ratio of the coupling under which $\chi_\phi$ is charged to the dominant coupling $g$.
We restrict ourselves to order of magnitude estimation
and do not care about this factor in the following discussion
to avoid complications.
}
\begin{align}
	g^2 (\phi) \left( \int_{\bm p} \frac{f_\chi (\bm{p})}{p} \right)^2 \equiv
	g^2 (\phi) T_{\ast}^4.
\end{align}
Here we have defied the effective temperature $T_\ast$ for light particles $\chi$,
which does not always coincides with that for $\chi_\phi$-particles since they
have masses of $|g_\phi \phi|$.
Extracting $\phi$-dependent part, one may obtain
\begin{align}
	 V_\text{L} \sim \alpha^2 T_\ast^4 \log \lmk \frac{\phi^2}{T_\ast^2} \rmk,
	 \label{eq:log}
\end{align}
with $\alpha = g^2 / 4 \pi$.
This is nothing but the so-called thermal-log potential for the thermal distribution of $\chi$-particles~\cite{Anisimov:2000wx}.

Thus, in order to discuss the finite density corrections to the effective potential for $\phi$,
we have to know distribution functions for 
light $\chi$-particles 
as well as that for 
$\chi_\phi$-particles with the mass of $|g_\phi \phi|$. 
In the following, we discuss how $\chi$ and $\chi_\phi$ are produced from hard primaries
during the course of thermalization
and calculate their effective temperatures $T_\ast$ and $T_{\ast, \chi_\phi}$.

Before closing this section,
we would like to comment on basic assumptions behind our following discussions.
First of all,
the effective potential listed above is computed perturbatively,
which implicitly assumes that the occupation number $f(k_\text{max})$ 
[that dominates the integral $\int_{\bm k} f/k \sim f(k_\text{max}) k_\text{max}^2$]
should be smaller than $\alpha^{-1}$, 
since otherwise the perturbative expansion breaks down.
Below, we will see that this condition is automatically satisfied.
In addition,
although we only focus on the effective potential to discuss the symmetry restoration,
it is not trivial that whether or not the effective potential is maintained after the onset of oscillation
and also the scalar condensate soon dissipates its energy into background plasma.
In the following,
we simply assume that $\phi$ can dissipate its energy into background plasma
soon after the onset of its oscillation,
since its relaxation process can be strongly model dependent~\cite{Moroi:2013tea}.

\section{\label{thermalization}Thermalization History}

In this section, we investigate the thermalization process of hard primaries 
and calculate the effective temperatures for the soft sector $T_\ast$ and $T_{\ast, \chi_\phi}$. 
Basic parameters which are required to study thermalization after inflation
are the mass of inflaton $m_I$, the decay rate of inflaton $\Gphi$
and the dominant couplings of decay produces $\alpha$;
{\it e.g.,} for the SM plasma, the predominant interaction is the strong coupling, $\alpha = \alpha_s$.
This is because the first two parameters $m_I$, $\Gphi$ determine the typical distribution of decay products,
and the last parameter $\alpha$ is essential to estimate their interaction time scale.
We are also interested in the production of $\chi_\phi$-particles
and hence the coupling between $\chi_\phi$ and the scalar condensate $\phi$,
which is given by $g_\phi$,
is also an important parameter.
Thus, we have four parameters essentially to capture this system.

We assume that the inflaton decays into particles $\chi_I$ 
which are charged under the abelian and/or non-abelian gauge groups (SM gauge group). 
We mainly consider the era after $\ti = \ti_{\rm ini}$, 
where $\ti_{\rm ini}$ is defined below [see Eq.~\eqref{eq:initial}]. 
First, 
the hard primaries emit soft particles 
but the soft sector is not thermalized soon. 
As shown in Sec.~\ref{phase2}, 
the soft sector is completely thermalized 
after the time of $\ti = \ti_{\rm soft}$ [Eq.~\eqref{eq:soft}]. 
After that, around the time of $\ti = \ti_{\rm max}$ [Eq.~\eqref{eq:max}], 
the hard sector as well as the soft sector are completely thermalized 
and the effective temperature reaches a (local) maximum. 
Then the reheating is completed at $\ti = \ti_{\rm RH}$ [Eq.~\eqref{eq:RH}]. 
We investigate thermalization processes during these eras in turn, 
and then we briefly summarize the result of this section in Sec.~\ref{summary in sec 3}. 
Our procedure is based on the one used in Ref.~\cite{Kurkela:2011ti,Kurkela:2014tea} (see also Refs.~\cite{Harigaya:2013vwa, Harigaya:2014waa}).

Figs.~\ref{distribution1}, \ref{distribution2} and \ref{distribution3}
summarize the essence of this section.
Hurry readers can skip all the details and move to these figures.

\subsubsection*{Kinetic equations}

Here we briefly summarize minimal knowledge required for our estimation.
Eventually, we will see that the collinear splitting process in medium plays
crucial roles in the following discussion.
We also summarize basic formulae of this process in SU$(N)$ with $N_F$-flavor fermions
in Sec.~\ref{app:lpm} for the sake of completeness.

Our method is based on  kinetic equation (Boltzmann-like equation),
which can be derived from first principles, {\it i.e.},~from 
Schwinger Dyson equations in the closed time path formalism
(see for instance \cite{baym1962quantum, Calzetta:1986cq},
reviews \cite{Blaizot:2001nr,Berges:2004yj} and references therein).
It was shown that particle-like excitations of $m_s \ll  p$ with $m_s$ being the screening mass
obey the kinetic equations if these quasi-particles fulfill the following conditions:
(i) Typical ``size'' of quasi-particles are smaller than the mean free path,
(ii) Typical duration time of interactions are shorter than the mean free time,
(iii) Typical value of distribution function should be smaller than $1/\alpha$.
The condition (ii) is sometimes violated for collinear emissions in the medium,
and in this case appropriate resummations are required, which is so-called the
Landau-Pomeranchuk-Migdal (LPM) effect~\cite{Landau:1953um, Migdal:1956tc,Gyulassy:1993hr,Arnold:2001ba,Arnold:2001ms, Arnold:2002ja,Besak:2010fb}.
After the resummations, we eventually obtain the following kinetic equations for quasi-particles
in the plasma~\cite{Arnold:2002zm,Kurkela:2011ti,Kurkela:2014tea}:
\begin{align}
	\left[ \partial_t - H \bm{k} \cdot \frac{\partial}{\partial \bm{k}} \right]
	f_\bullet (t, k) = - {\cal C} [f_\bullet],
\end{align}
where $f_\bullet$ is the distribution function of quasi particles with $\bullet$ being species
and ${\cal C}$ is the collision term.
As we will explain later,
at least for the perturbative Planck-suppressed decay of inflaton,
these conditions are safely satisfied\footnote{
	It is contrary to the preheating case where the fluctuations can be strongly correlated.
	See Refs. \cite{Micha:2004bv, Berges:2008wm} for instance.
}
except for the condition (ii).
Hence, we can rely on the kinetic equations given in \cite{Arnold:2002zm,Kurkela:2011ti,Kurkela:2014tea}.
In our case, we are interested in the distributions of $\chi$ and $\chi_\phi$
to obtain information of effective temperatures.

The collision term for $\chi$-particles encodes two-to-two scatterings responsible for diffusions
and effective ``one-to-two'' (inelastic) scatterings due to splittings:
\begin{align}
	{\cal C} [f_{\chi}] \supset 
	\left( {\cal C}_\text{2 to 2} + {\cal C}_\text{split} \right) [f_{\chi}],
	\label{eq:collision}
\end{align}
where
\begin{align}
	{\cal C}_\text{2 to 2} [f_a]
	= \frac{1}{2\nu_a} \sum_{b,c,d} \int_{\bm{p, k', p'}} 
	&\frac{\left| {\cal M}^{ab}_{cd} (K, P ; K' , P') \right|^2}{2 E_k 2 E_p 2 E_{k'} 2 E_{p'}}
	\left( 2 \pi \right)^4 \delta^{(4)} \left( K + P - K' - P' \right) \nonumber\\
	\times & \com{
		f_a (k) f_b (p)  \left[ 1 \pm f_c (k') \right]  \left[ 1 \pm f_d (p') \right]
		- \left( \text{inverse process} \right)
	},
\end{align}
\begin{align}
	{\cal C}_\text{split} [f_a] = & \quad
	\frac{\prn{2 \pi}^3}{2 k^2 \nu_a} \sum_{b,c}
	\int \dd p' \dd k' \delta \prn{k - p' - k'} \gamma^a_{bc} \prn{k; p', k'} \nonumber \\
	&\quad \quad \quad \quad \quad  
	\times \com{
		f_a (k) \com{ 1 \pm f_b (p') } \com{ 1 \pm f_c (k') } - 
		\left( \text{inverse process} \right)
	} \nonumber \\[.5em]
	&+ \frac{\prn{2 \pi}^3}{k^2 \nu_a} \sum_{b,c}
	\int \dd p' \dd p \delta \prn{k + p - p'} \gamma^c_{ab} \prn{p'; k, p} \nonumber \\
	&\quad \quad \quad \quad \quad  
	\times \com{
		f_a (k) f_b (p) \com{ 1 \pm f_c (p') } - 
		\left( \text{inverse process} \right)
	} \\[1em]
	\sim & \quad
	\int \dd \ln p' \, \Gamma_\text{split} (p') 
	\com{
		f_\chi (k) \com{ 1 \pm f_\bullet (k - p') } \com{ 1 \pm f_\bullet (p') }
		-\left( \text{inverse process} \right)
	}
	\nonumber \\
	& +\int \dd \ln p' \, \prn{ \frac{p'}{k} }^3
	\Gamma_\text{split} (k)
	\left[ f_\chi (k) f_\bullet (p'-k) \left[ 1 \pm  f_\bullet (p') \right] -
	\left( \text{inverse process} \right) \right],
	\label{eq:split_apprx}
\end{align}
with $\nu_a$ being the number of degree of freedom for a species $a$
(normalized by one real d.o.f.).
Here, the signs are taken as $+$ ($-$) for boson fields (fermion fields). 
$\Gamma_\text{split}$ stands for the splitting rate.
One can show that Eq.~\eqref{eq:split_apprx} can be derived by using 
the definition of the splitting function, $\gamma^a_{bc}$, for $a \leftrightarrow bc$,
given in App.~\ref{app:lpm}.
Instead, below, we give an intuitive derivation of the splitting rate, $\Gamma_\text{split}$,
in order to understand its physics.
See App.~\ref{app:lpm} for more rigorous explanation.

\begin{figure}[t]
\centering 
\includegraphics[width=.45\textwidth
]{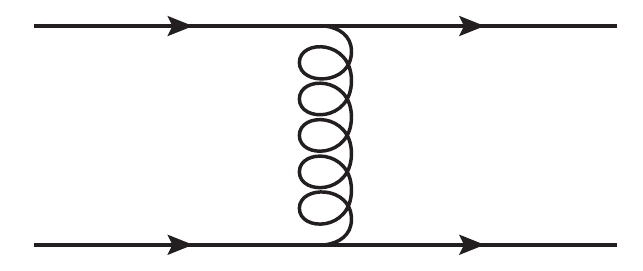} 
\caption{\small
Diagram for elastic 2 to 2 scattering process. 
}
  \label{fig:diag3}
\end{figure}

The first term of the right-hand side of Eq.~(\ref{eq:collision}) imprints elastic scatterings that diffuse distribution functions in momentum space, where
the process is dominated by small momentum exchange due to the strong $t$-channel enhancement of gauge bosons 
as shown in Fig.~\ref{fig:diag3}. 
The momentum transfer squared obeys the following diffusion equation:
\begin{align}
	\left( \Delta p \right)^2 \sim \qel t,
\end{align}
with
\begin{align}
	\qel \sim \int \dd^2 q_\perp \frac{\del \Gamma_{\rm el}}{\del q_\perp^2} q_\perp^2 \sim \alpha^2 
	\int_{p'} f_\bullet (p') \lkk 1 \pm f_\bullet (p') \rkk. 
	\label{qel}
\end{align}
Here we have used the rate of elastic scatterings
\begin{align}
	 \frac{\del \Gamma_{\rm el}}{\del q_\perp^2} \sim
	 \frac{\alpha^2}{q_\perp^2 ( q_\perp^2 + m_s^2)} 
	 \int_{\bm p'} f_\bullet (p') \lkk 1 \pm f_\bullet (p') \rkk.
	 \label{Gamma_el}
\end{align}
Note that the screening mass for $\chi$-particles is given by $m_s^2 \sim \alpha T_\ast^2$.

\begin{figure}[t]
\centering 
\includegraphics[width=.95\textwidth
]{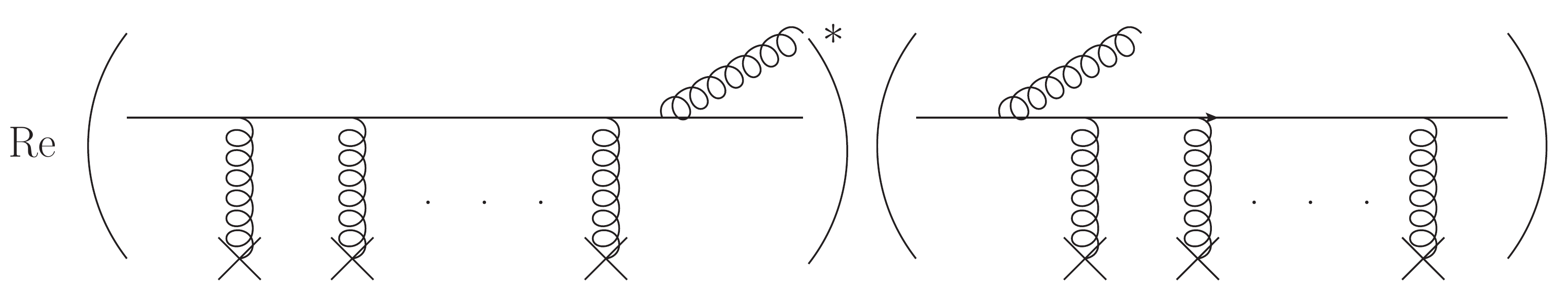} 
\caption{\small
Diagram representing an interference effect, known as the LPM effect. 
}
  \label{fig:diag4}
\end{figure}

Then, let us move to the second term, ${\cal C}_\text{split}$.
As notified in the beginning,
in estimating the inelastic scattering rate of the hard primaries, it is necessary to
take into account an interference effect between a hard primary 
and its daughter particle, 
known as the LPM effect.
A Feynman diagram describing the interference effect is shown in Fig.~\ref{fig:diag4}. 
The interference effect forbids subsequent scattering processes 
during the interval while their phase spaces overlap with each other. 
Taking this effect into account, 
we obtain the following rate of inelastic scatterings: 
\begin{align}
	 \Gamma_\text{split} (k) \sim \alpha \ \Min \lkk \Gamma_{\rm el}, \  \frac{1}{t_\text{form}} \rkk, 
	 \label{splitting rate}
\end{align}
with
\begin{align}
	t_\text{form} \equiv \sqrt{\frac{k}{\qel}}.
	\label{eq:t_form}
\end{align}
Here we omit a factor that comes from model dependence and logarithmic enhancement.
See App.~\ref{app:lpm}.
The time scale of $t_\text{form}$ denotes the formation time which it takes to resolve
the overlap between the daughter and parent particles.
$k$ is the momentum for a daughter particle. 
One can see that the rate is LPM suppressed above the threshold momentum $k_\text{LPM}$,
which is given by
\begin{align}
	k_\text{LPM} \equiv \frac{\qel}{\Gamma_\text{el}^2}.
\end{align}
Note that Eq.~\eqref{eq:t_form} comes from the condition so that the overlap between
the parent and daughter should be resolved; $t \gtrsim k / k_\perp^2$ with $k_\perp^2 \sim \hat q_\text{el} t$.
Therefore, for a given time $t$, there is an upper bound on the daughter momentum:
\begin{align}
	k_\text{form} = \hat q_\text{el} t^2.
\end{align}
Above $k_\text{form}$, even the diffusion cannot resolve the destructive interference.
In this case, the vacuum cascades may become important,
though the emitting angle between the parent and the daughter is bounded below
$\theta \gtrsim (kt)^{-1/2}$ for a given $k$ and $t$.
See also Eq.~\eqref{f_s before t_ini} and discussion about it.

\begin{figure}[t]
\centering 
\includegraphics[width=.45\textwidth
]{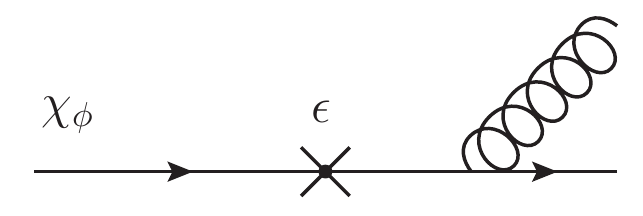} 
\caption{\small
Diagram for the decay of $\chi_\phi$-particles. 
}
  \label{fig:diag5}
\end{figure}

So as to obtain the effective temperature of $\chi_\phi$,
we also have to know how $\chi_\phi$-particles are produced.
The collision term for $\chi$-particles responsible for $\chi_\phi$-production
may be given by
\begin{align}
	{\cal C} [f_{\chi_\phi}] \supset 
	\left( {\cal C}_\text{2 to 2} + {\cal C}_\text{split} + {\cal C}_\text{dec} \right) [f_{\chi_\phi}],
\end{align}
where the last term with the subscript ``dec'' denotes the decay/inverse decay processes. 
They are expected in general since the $\phi$-condensation can give a sizable mass to $\chi_\phi$-particles;
that is, larger than the screening mass, $|g_\phi \phi| > m_s$, 
and are calculated by 
\begin{align}
	{\cal C}_\text{dec} [f_{\chi_\phi}]
	\sim \int_{\bm{k', p'}} 
	&\frac{\left| {\cal M}_\text{dec} (K; K' , P') \right|^2}{2 E_k 2 E_{k'} 2 E_{p'}}
	\left( 2 \pi \right)^4 \delta^{(4)} \left( K - K' - P' \right) \nonumber\\
	&\left[
		f_{\chi_\phi} (k) \left[ 1 \pm f_\bullet (k') \right]  \left[ 1 \pm f_\bullet (p') \right]
		-
		\left( \text{inverse process} \right)
	\right].
\end{align}
These collision terms depend on models how $\chi_\phi$-particles couple with other light particles;
in particular, we have to specify interaction terms which can induce decay of $\chi_\phi$ into light particles
for a sizable mass of $\chi_\phi$, $|g_\phi \phi| > m_s$.
For simplicity, we assume that the typical magnitude of this term is proportional to
$\epsilon^2 \alpha$ with $\epsilon$ being a small parameter 
and $\alpha$ being the fine structure constant of the gauge group which dominates the thermalization.
We assume $\epsilon^2 \lesssim \alpha$ to avoid unnecessary complications.
See also discussion given below Eq.~\eqref{eq:inv_decay}.
For instance, if $\chi_\phi$-particles are charged under the gauge group and if
they mix with light particles, then their decay rate may be proportional to $\epsilon^2 \alpha |g_\phi \phi|$,
where $\epsilon$ is identified as the mixing angle 
(see Fig.~\ref{fig:diag5}). 
Under this assumption, the dominant contribution to first two terms is essentially the same as that for $\chi$-particles,
though we should note that 
when a daughter particle has a mass larger than the screening mass $m_s$, 
we have to replace $m_s$ 
in the scattering rate of Eq.~(\ref{Gamma_el}) to the mass of the daughter particle.
And also note that these terms can be sources of $\chi_\phi$-particle production.

\subsection{
$\ti_\text{ini} < \ti < \ti_{\rm soft}$
}
\label{sec:phase1}

Apparently, two to three processes with the momentum exchange of the order of $m_I$
seem to be efficient in order to decrease/increase energy/number of hard primaries.
However, this is not true since the reaction energy/occupation number is too large/small
and the rate is found to be much smaller than the Hubble parameter.
This is a typical situation of reheating via Planck suppressed decay, $\tilde \Gamma_I \lesssim  1$.
In this case, the hard primaries lose their energy mainly through inelastic collinear scatterings 
with the rate of Eq.~(\ref{splitting rate}), which are enhanced by the $t$-channel contribution~\cite{Davidson:2000er}. 
Below, we will see how $\chi$-particles are produced via inelastic scatterings,
and form the soft population.
Then, we discuss production of $\chi_\phi$-particles, which is essentially different from
that of $\chi$-particles due to the mass of $|g_\phi \phi|$.

\subsubsection*{Production of $\chi$-particles}
At the first stage of thermalization process, 
the hard primaries scatter among themselves and emit soft particles 
through the inelastic scatterings imprinted in ${\cal C}_\text{split}$. 
The phase space distribution of these soft particles might be estimated as 
\begin{align}
	 f_s (t, k) \sim \Gamma_\text{split} (k) n_h k^{-3} t, 
	 \label{f_s}
\end{align}
where $n_h \sim \int_{\bm p} f_h$ is the number density of hard primaries [see Eq.~(\ref{hard-distribution})]. 
However, low-momentum modes are so over-occupied that 
inverse processes and subsequent elastic scattering processes have to be taken into account. 
In fact, soft particles feel subsequent scatterings with hard primaries\footnote{
	Although there are many processes for the thermalization of soft modes (see Ref.~\cite{Kurkela:2011ti,Kurkela:2014tea}), 
	we explain one of them as an illustration.
} 
and 
low-momentum modes fall into a thermal-like distribution 
such as\footnote{
	In the case that soft particles are fermions, 
	the distribution $f_s$ cannot exceed unity due to the Pauli blocking effect. 
	In this case, we should truncate $f_s$ when it reaches unity. 
	However, we expect that 
	soft particles contain bosons that dominate the thermal effect on $\phi$, 
	so that we focus on the distribution of bosonic particles. 
	Its extension to fermionic case is straightforward.
}
\begin{align}
	f_s (t, k) \sim 
	\frac{T_s }{ k }
	 \label{f_s thermal 2}
	 ~~\text{for}~~k < k_{\rm max}, 
\end{align}
where $k_{\rm max}$ is 
determined by the diffusion constant of Eq.~(\ref{qel}) with $f = f_h$ 
and is written as
\begin{align}
	 k_{\rm max} \sim \sqrt{\qel t} \sim \alpha \mphi \sqrt{f_h (t,m_I) \ti}
	 \sim \alpha m_I \Gphi^{1/2}.
	 \label{k_max 2}
\end{align}
One can show that it is the same as the threshold momentum of LPM suppression, 
i.e., $k_\text{max} \sim k_\text{LPM}$.
The effective temperature for bosonic soft modes $T_s$ can be estimated from 
the energy conservation. 
The energy of soft modes below the scale $k_{\rm max}$ is 
given as\footnote{
	One might wonder if the energy conservation within the soft modes 
	[that is, Eq.~(\ref{E conservation})] is violated 
	due to the scatterings between hard primaries and soft modes. 
	However, this does not change our order estimations. 
	The energy for soft modes is dominated by that for the modes around $k_{\rm max}$, 
	and the scatterings between hard primaries and such soft modes 
	are inefficient to change the energy-conservation relation of Eq.~(\ref{E conservation}). 
}
\begin{align}
	 k_{\rm max} \Gamma_\text{split} (k_{\rm max}) t n_h \sim 
	 \rho_s \sim k_{\rm max} n_s \sim T_s k_{\rm max}^{3}, 
	 \label{E conservation}
\end{align}
where $\Gamma_\text{split}$ is given by Eq.~(\ref{splitting rate}). 
Thus, the effective temperature for the soft modes $T_s$ is given as 
\begin{align}
	T_s \sim \alpha^{-1/2} \Gphi^{1/4} \ti^{-1/2} \mphi, 
\end{align}
from Eq.~\eqref{E conservation} together with Eqs.~\eqref{splitting rate} and \eqref{k_max 2}. 
Here, we have used $\Gamma_\text{split} \sim \alpha \sqrt{\qel/k}$ 
because the relevant energy scale $k_{\rm max}$ is so large that 
such a soft mode are produced by LPM suppressed interactions.

\begin{figure}[t]
\centering 
\includegraphics[width=.7\textwidth
]{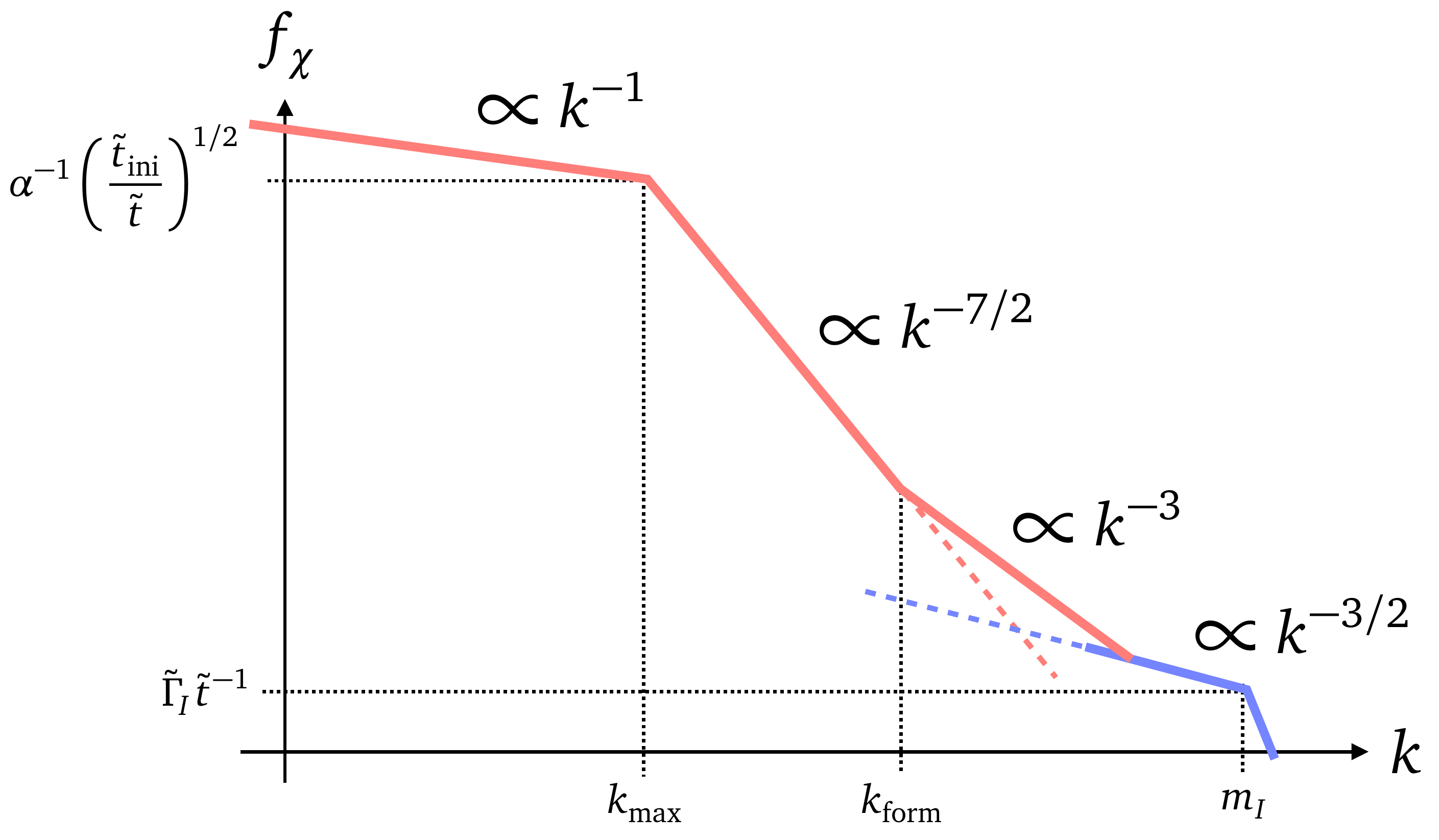} 
\caption{\small
Distribution of $\chi$-particles for $\ti_{\rm ini} < \ti < \ti_{\rm soft}$. 
Recall that $ k_\text{max} \sim \alpha \tilde \Gamma_I^{1/2} m_\phi$,
$k_\text{form} \sim k_\text{max} (\tilde t / \tilde t_\text{ini})$
and $t_\text{ini} \equiv \alpha^{-1} \tilde \Gamma_I^{-1/2}$.
{\bf Soft} ({\color{ppink}pink}):
The thermal-like distribution dominates below $k_\text{max}$,
the LPM-suppressed spectrum can be seen between $k_\text{max}$ and $ k_\text{form}$,
and the vacuum cascade may be relevant above $k_\text{form}$.
{\bf Hard} ({\color{bblue}blue}):
The hard spectrum is sourced by the direct decay of inflaton and
by its red-shifted spectrum of previously produced one.
See Eq.~\eqref{hard-distribution}.}
  \label{distribution1}
\end{figure}

We illustrate the distribution of $\chi$-particles in Fig.~\ref{distribution1}, 
where we include the hard sector as well as the soft one. 
The distribution of smaller wavenumber modes than $k_{\rm max}$ 
is a thermal-like one given by Eq.~(\ref{f_s thermal 2}), 
while that of larger wavenumber modes in the soft sector 
is determined by the LPM effect as Eq.~(\ref{f_s}). 
The distribution of hard sector is given by Eq.~(\ref{hard-distribution}). 
The distributions of soft and hard sectors coincide with each other 
at a certain wavenumber, 
$\Gphi^{-1/4} k_\text{max} \ti^{1/4}$, 
shown in Fig.~\ref{distribution1}. 
For clarity, we call the distribution of $\chi$-particles above/below this momentum as the hard/soft sector.

At this regime, the effective temperature $T_*$ is dominated by the contribution from the soft sector
\begin{align}
	 T_*^2 &\sim \int_{\bm k} \frac{f_s}{k} 
	 \sim T_s k_{\rm max} 
	 \sim \alpha^{1/2} \Gphi^{3/4} \ti^{-1/2} \mphi^2, \label{eq:efft_tini_tsoft} \\
	& > 
	 \int_{\bm k} \frac{f_h}{k} \sim \Gphi \ti^{-1} m_I^2, 
	 \label{eq:effT_softvshard}
\end{align}
where the inequality holds for $\ti > \ti_{\rm ini} \equiv \alpha^{-1} \Gphi^{-1/2}$ (see below). 
This implies that the screening mass $m_s$ ($= \alpha T_*^2$) is already dominated by the soft sector.
Note that 
$T_s \ne T_*$ 
because the soft sector is not completely thermalized and 
$T_s$ and $T_*$ are nothing but effective temperatures with different definitions.

Here, we should ensure that the relevant wavelength scale $({\it e.g.},~k_{\rm max}^{-1})$ is smaller than 
the Hubble horizon $H^{-1} \sim t$ for a consistent treatment
because 
longer wavelength modes than the Hubble horizon cannot be emerged due to the loss of causality. 
This implies that 
we can determine the time scale when the softest mode emerges 
by comparing the maximum momentum of the soft sector with the Hubble parameter. 
The inequality $k_\text{max} > H$ indicates
\begin{align}
	\ti > \ti_\text{ini} \equiv \alpha^{-1} \Gphi^{-1/2}.
	\label{eq:initial}
\end{align}
This inequality immediately means that the occupation number of soft sector at $k_\text{max}$,
which dominates the energy and number density of the soft sector,
is smaller than $\alpha^{-1}$:
\begin{align}
	f_s (t, k_\text{max}) \sim \frac{T_s}{k_\text{max}} 
	\sim \alpha^{-1} \lmk \frac{\ti_{\rm ini}}{t} \rmk^{1/2} 
	< \alpha^{-1}.
	\label{eq:perturbative}
\end{align}
Hence, we may treat the soft sector perturbatively.
Then, let us compare the screening mass with the maximum momentum $k_\text{max}$.
Up to here, our arguments rely on the effective kinetic theory of non-Abelian plasma,
and hence produced modes 
should not be too soft, $k > m_s$, for a consistent treatment.
In fact, Eq.~\eqref{eq:initial} also ensures
\begin{align}
	\frac{k_\text{max}}{m_s} \sim \left( \frac{\ti}{\ti_\text{ini}} \right)^{1/4} > 1.
	\label{eq:consistent}
\end{align}
We should also note that 
\begin{align}
	 \frac{m_s}{H} \sim \lmk \frac{\ti}{\ti_{\rm ini}} \rmk^{3/4} > 1, 
\end{align}
which means that the screening length is smaller than the horizon length. 
Finally, we check that 
a typical time scale of interactions ({\it e.g.,} $\Gamma_{\rm el}^{-1}$) 
is smaller than the cosmological time scale $H^{-1}$: 
\begin{align}
	 \frac{H^{-1}}{\Gamma_{\rm el}^{-1}} 
	 \sim \lmk \frac{\ti}{\ti_{\rm ini}} \rmk^{1/2} > 1, 
\end{align}
where we use Eq.~(\ref{Gamma_el}). 
Therefore, 
the condition $\ti > \ti_{\rm ini}$ 
ensures that 
$k_{\rm max} > m_s > H$, $f_s < \alpha^{-1}$, 
and $\Gamma_{\rm el} > H$. 
These inequalities justify our calculations performed in this paper. 
Otherwise we may have to take into account the effect of causality, 
the finite cosmological time scale, 
and non-perturbative effects on production processes.

Finally, we comment on the era before $\ti_\text{ini}$.
Since elastic scatterings among decay products does not take place within the Hubble time scale,
medium induced cascades can be neglected.
In addition, the formation momentum, $k_\text{form}$, is smaller than the Hubble parameter
before $t_\text{ini}$:
\begin{align}
	\frac{k_\text{form}}{H} \sim \prn{ \frac{\tilde t}{\tilde t_\text{ini}} }^2 < 1.
\end{align}

Still, the vacuum cascades might broaden the spectrum towards the infrared,
and hence we roughly estimate their effect on the effective temperature.
The quantum formation time scale in vacuum implies that one can find a quanta of $k$ with a probability of
${\cal O} (\alpha)$ per $\ln k \ln \theta$ for $\theta > (k t)^{-1/2}$.
Omitting the log factor, one may obtain the distribution of the soft sector and the effective temperature as
\begin{align}
	f_s (t, k) \sim \alpha \Gphi \ti^{-1} \left( \frac{m_I}{k} \right)^3 ~~ \text{for}~~k > H
	\to
	T_\ast^2 \sim \alpha \Gphi m_I^2,
	\label{f_s before t_ini}
\end{align}
which coincides with Eq.~\eqref{eq:efft_tini_tsoft} for $\ti \to \ti_\text{ini}$.
Hence, we expect that the effective temperature is at most that of $\ti_\text{ini}$.
After $\ti_\text{ini}$, these particles with $k^{-3}$ are soon red-shifted away
and LPM suppressed spectrum starts to dominate the soft sector.
Thus, we will not consider the case of $\ti < \ti_\text{ini}$ further in the following.

\subsubsection*{Production of $\chi_\phi$-particles}

In contrast to $\chi$-particles, $\chi_\phi$-particles can have sizable masses of $|g_\phi \phi|$
due to the large expectation value of the $\phi$-condensate.
The distribution in momentum space can be different from that of $\chi$-particles and hence separate discussions are required.
Although the calculations are complicated, 
we find that 
the effect on the effective potential from $\chi_\phi$-particles 
is always subdominant for the case of $|g_\phi \phi | \gtrsim \alpha^{-1} k_{\rm max}$ and $\Gphi \lesssim \alpha^{6}$. 
Even in the case of $\Gphi \gtrsim \alpha^{6}$, its effect is roughly the same order with 
that of thermal log potential from $\chi$-particles, 
so that we can neglect it for rough estimations. 
We briefly explain $\chi_\phi$-particle production processes in this subsection, 
while detailed calculations are shown in Appendix.\footnote{
	We conservatively omit the decay of massive $\chi_\phi$-particles 
	to show that the effect on the effective potential from $\chi_\phi$-particles 
	is usually subdominant. 
	In other words, 
	we show that the effect from $\chi_\phi$-particles can be neglected 
	even if they are stable and maximally abundant. 
	Note that the calculation can be applied to the production of massive dark matter, 
	once we forget the field $\phi$ and identify $\chi_\phi$ as just a dark matter with mass of $|g_\phi \phi|$. 
	Its abundance can be calculated from 
	the resulting distribution function of $\chi_\phi$-particles given in Appendix. 
	\label{footnote:decay}
}

If the effective mass, $|g_\phi \phi|$, is smaller than the screening mass, $m_s$,
then the results are exactly the same as those for $\chi$-particles.
Hence, we concentrate on $|g _\phi \phi| > m_s$ in the following.
First, let us consider contributions from splittings which are imprinted in ${\cal C}_\text{split}$.
Note that since we consider the case of $|g_\phi \phi | > m_s$, 
we should replace $m_s$ in Eq.~(\ref{Gamma_el}) to $|g_\phi \phi |$. 
The phase space distribution and the resultant effective temperature may be estimated as 
\begin{align}
	\left. f_{\chi_\phi} (t, k) \right|_\text{hard} \sim
	\Gamma_\text{split} (k) n_h k^{-3} t
	~\to ~
	\left. T_{\ast, \chi_\phi}^2 \right|_\text{hard}
	\sim \Gamma_\text{split} (M) n_h M^{-1} t,
\end{align}
where
\begin{align}
	M \equiv \Max \left[ |g_\phi \phi|, k_{\text{max}, \chi_\phi} \right];~~
	k_{\text{max}, \chi_\phi} \equiv k_\text{max} \left( \frac{m_s}{ | g_\phi \phi |} \right).
\end{align}
The ``hard'' subscript means that splittings occur dominantly through interactions among hard primaries,
since the number density is still dominated by the hard sector at this stage, $\ti < \ti_\text{soft}$.
Here $\Gamma_\text{split}$ is given by Eq.~(\ref{splitting rate}) but $m_s$ is replaced by $|g_\phi \phi |$. 
In this case, the threshold momentum $k_{\text{LPM}, \chi_\phi}$ is given by
\begin{align}
	k_{\text{LPM}, \chi_\phi} = k_\text{LPM} \left( \frac{| g_\phi \phi |}{ m_s} \right)^4.
\end{align}
If the effective mass for $\chi_\phi$ is not so large, that is, $( g_\phi \phi )^2 < m_s k_{\text{max}, \chi_\phi}$, 
then the distribution function for $\chi_\phi$ coincides with that for $\chi$ below $k_{\text{max}, \chi_\phi}$.

In addition, $\chi_\phi$-particles are produced via ${\cal C}_\text{2 to 2}$ and ${\cal C}_\text{dec}$ with 
interactions among the soft sector and also between the soft and hard sector.
First, let us concentrate on ${\cal C}_\text{2 to 2}$,
since it gives model independent contribution.\footnote{
	We can drop terms proportional to $\epsilon^2 \alpha^2$ at least for $\epsilon^2 \lesssim \alpha$.
}
Although the number density is still dominated by the hard sector,
the soft sector may produce ``soft'' $\chi_\phi$-particles with a larger rate ({\it e.g.}, via $s$-channel scattering processes). 
The 
effective temperature may be given by
\begin{align}
	\left. T_{\ast, \chi_\phi}^2 \right|_{\rm soft}
	&\sim  
	\frac{\alpha^2}{| g_\phi \phi |^3} t
	\int \dd \log k' \ 
	n_s (k') n_s \left( \frac{|g_\phi \phi|^2}{k'}  \right), 
	\label{eq:teff_soft_phase1}
\end{align}
for $|g_\phi \phi| > k_\text{max}$, where $n_s \sim \int_{\bm p} f_s$.
Note that, for $|g_\phi \phi| < k_\text{max}$, it is given by $\left. T_{\ast, \chi_\phi}^2 \right|_{\rm soft} \sim k_\text{max}^2$.
They are also produced from interactions between the hard and soft sectors: 
\begin{align}
	\left. T_{\ast, \chi_\phi}^2 \right|_{\rm int}
	&\sim  
	\frac{\alpha^2}{| g_\phi \phi |^3} t
	\int \dd \log k' \ 
	n_s (k') n_h \left( \frac{| g_\phi \phi |^2}{k'}  \right), 
	\label{eq:teff_int_phase1}
\end{align}
where the subscript ``int'' indicates contributions from the interaction between the hard and soft sectors. 
Note that both $n_s$ and $n_h$ can be calculated from the distribution function 
illustrated in Fig.~\ref{distribution1}. 
Then, we give comments on contributions from ${\cal C}_\text{dec}$,
which strongly depends on details of a model.
As an illustration, we have assumed that the magnitude of this process is proportional to $\epsilon^2 \alpha$.
The inverse decay can produce $\chi_\phi$-particles and it may yield
\begin{align}
	\left. T_{\ast, \chi_\phi}^2 \right|_{\rm soft/int}
	\sim 
	\frac{\epsilon^2 \alpha}{|g_\phi \phi|^3} t
	\int \dd \log k' n_s (k') n_{s/h} \left(\frac{|g_\phi \phi|^2}{k'} \right).
	\label{eq:inv_decay}
\end{align}
Hence, if the typical mixing parameter $\epsilon$ is not so large, say $\epsilon^2 \lesssim \alpha$,
then this contribution does not exceed those from Eqs.~\eqref{eq:teff_soft_phase1} and 
\eqref{eq:teff_int_phase1}.
Moreover, this collision term, ${\cal C}_\text{dec}$, also yields decay of $\chi_\phi$-particles:
\begin{align}
	\vev{ \Gamma_\text{decay} } t \sim \epsilon^2 \alpha |g_\phi \phi| t \lesssim
	\left( \frac{|g_\phi \phi|}{k_\text{max}}  \right) \left( \frac{\ti}{\ti_\text{soft}} \right), 
\end{align}
where we use $\ti_{\rm soft} \equiv \alpha^{-3} \Gphi^{-1/2}$ defined in the next subsection. 
One can see that the decay is insignificant for $| g_\phi \phi | < k_\text{max}$ at this stage, $\ti \lesssim \ti_\text{soft}$.
As explained in the footnote~\ref{footnote:decay},
we can omit the decay term in order to show that the effect on the effective potential from abundant $\chi_\phi$-particles
is subdominant for $|g_\phi \phi| \gtrsim \alpha^{-1} k_\text{max}$.

For later convenience,
we denote all these contributions where $\chi_\phi$-particles are produced indirectly (not directly produced by inflaton decay) as 
\begin{align}
		\left. T_{\ast, \chi_\phi}^2 \right|_\text{indir} \equiv \max \left[  
		\left. T_{\ast, \chi_\phi}^2 \right|_\text{hard},
		\left. T_{\ast, \chi_\phi}^2 \right|_\text{soft},
		\left. T_{\ast, \chi_\phi}^2 \right|_\text{int}
		\right].
\end{align}
The explicit forms of the RHS are shown in Appendix.

In contrast to the above discussion,
if $\chi_\phi$-particles are produced directly from inflaton decay,
then the situation turns out to be slightly different.
This is because the hard sector still has a sizable contribution
to the effective temperature,
which tend to dominate for a large field value. 
In this case, its distribution function has a contribution of Eq.~(\ref{hard-distribution}). 
As shown below, 
the effect on the effective potential from $\chi_\phi$-particles 
can be efficient compared with the thermal log potential. 
Therefore, in this case,
we should take into account the decay of $\chi_\phi$-particles 
to obtain more realistic predictions 
because $\chi_\phi$-particles obtain effective masses of $|g_\phi \phi |$ 
and can decay into light particles in general (see footnote~\ref{footnote:decay}). 
Its decay effect can be taken into account 
by multiplying $\Min [ 1, \ (\Gamma_{\rm decay} t)^{-1} ]$ 
to its distribution of Eq.~(\ref{hard-distribution}), 
where $\Gamma_{\rm decay}$ is the decay rate of 
$\chi_\phi$-particles. 
The effective temperature may be estimated as 
\begin{align}
	\left. T_{\ast, \chi_\phi}^2 \right|_\text{dir}
	= \Gphi \ti^{-1} m_I^2 \,
	\Min \left[ 1, \frac{1}{\Gamma_\text{decay} t} \right],
\end{align}
where $\Gamma_\text{decay} \sim \epsilon^2 \alpha |g_\phi \phi|$.

\subsubsection*{Thermal potential}
Here we summarize the thermal potentials given in Eqs.~\eqref{eq:mass_2} and \eqref{eq:log},
and discuss which contribution dominates for each case. 
See also Appendix~\ref{app:eff_t}.
Note that for an interval between $k_\text{max} < | g_\phi \phi |  < \alpha^{-1} k_\text{max}$,
the effective temperature may depend on $\epsilon^2$.
We do not care about this small interval for our rough estimation.

\begin{itemize}
\item{Indirect $\chi_\phi$-production (produced from hard primaries/soft daughters):}
\begin{itemize}
\item
The simplest case is a large field value regime, $| g_\phi \phi | > \alpha^{-1} k_\text{max}$. 
In the case of $\Gphi \lesssim \alpha^6$, 
the effective mass from abundant $\chi_\phi$-particles always gives subdominant corrections to 
the effective potential, meanwhile the log contribution dominates:
\begin{align}
	\alpha^2 \frac{T_\ast^4}{\phi^2} \sim
	\alpha_\phi \alpha T_\ast^2 \left( \frac{\ti_\text{ini}}{\ti} \right)^{1/2} \left( \frac{k_\text{max}}{|g_\phi \phi|} \right)^{2}.
	\label{eq:log_p1_idir_l}
\end{align}
Even in the case of $\Gphi \gtrsim \alpha^6$, 
the effect of $\chi_\phi$-particles is very limited, 
so that we can neglect its contribution for a rough estimation.

\item
For a small field value regime, $| g_\phi \phi | < k_\text{max}$,
the effective mass from abundant $\chi_\phi$-particles dominates the effective potential:
\begin{align}
	\alpha_\phi \left. T_{\ast, \chi_\phi}^2 \right|_\text{indir} \sim
	\delta \alpha_\phi T_\ast^2;~~ \delta \in [\alpha, 1] .
\end{align}
Here we do not write down complicated parameter dependence of $T_{\ast, \chi_\phi} |_\text{indir}$ for brevity.
See Appendix~\ref{app:eff_t} for details.
\end{itemize}
\item{Direct $\chi_\phi$-production (produced from inflaton decay):}
\begin{itemize}
\item
For the large field value regime, $| g_\phi \phi | > \alpha^{-1} k_\text{max}$, 
the effective potential is given by the competition of two contributions;
the log dependence of the running coupling constant and hard $\chi_\phi$-primaries from the direct decay of inflaton.
The contribution from the direct decay of inflaton, which is given by
\begin{align}
	\left. T_{\ast, \chi_\phi}^2 \right|_\text{dir}
	= \alpha \Gphi^{1/2} T_\ast^2 \left( \frac{\ti_\text{soft}}{\ti} \right)^{1/2}
	\Min \left[ 1, \frac{1}{\Gamma_\text{decay} t} \right],
	\label{eq:p1_dir}
\end{align}
dominates the effective temperature for
\begin{align}
	m_I > |g_\phi \phi| > \alpha m_I \, \Max \left[ \alpha^{1/2} \Gphi^{1/4}, \alpha \left(\frac{\epsilon^2}{\alpha} \right) \left( \frac{\ti}{\ti_\text{soft}} \right) \right].
\end{align}
\item
For the small field value regime, $| g_\phi \phi | < k_\text{max}$, 
the contribution from the abundant $\chi_\phi$-particles dominates over that from the running coupling constant,
as explained in the previous case.
Again, we do not write down the explicit parameter dependence of $T_{\ast, \chi_\phi} |_\text{indir}$
since it is complicated. See Appendix~\ref{app:eff_t} for details.

The trivial case is for $|g_\phi \phi| < m_s$; there the contribution from abundant $\chi_\phi$-particles
always dominate over that from hard $\chi_\phi$:
\begin{align}
	\alpha_\phi \left. T_{\ast, \chi_\phi}^2  \right|_\text{indir} > 	\alpha_\phi \left. T_{\ast, \chi_\phi}^2  \right|_\text{dir}.
\end{align}
This is simply because the effective temperature for massless modes is dominated by the soft sector
as shown in Eq.~\eqref{eq:effT_softvshard}.

\end{itemize}
\end{itemize}

\subsection{$\ti_{\rm soft} < \ti < \ti_{\rm max}$ \label{phase2}}

After the time when the effective temperature $T_*$ becomes  comparable to 
$k_{\rm max}$, 
the soft sector is completely thermalized by their own interactions and can be described by a single parameter $T_s \sim T_\ast \sim k_\text{max}$. 
This occurs at the time of 
\begin{align} 
	 \ti \sim \ti_{\rm soft} \equiv \alpha^{-3} \Gphi^{-1/2}. 
	 \label{eq:soft}
\end{align}
At the same time, the number density of soft sector is as large as that of hard primaries, 
which means that we should substitute $f$ in Eq.~(\ref{qel}) with $f_s$.

\subsubsection*{Production of $\chi$-particles}
The energy injection to the soft sector is given by 
\begin{align}
	 \rho_s \sim k_{\rm split} n_h.
\end{align}
The momentum scale $k_{\rm split}$ is given by a criteria $\Gamma (k_{\rm split}) \sim H(t)$, 
which means that modes below $k_{\rm split}$ are soon thermalized and fall into soft modes. 
Equating $\rho_s$ with $T_s^4$, 
we obtain 
\begin{align}
	 T_s \sim \alpha^4 \Gphi \ti \mphi, \\
	 k_{\rm split} \sim \alpha^{16} \Gphi^3 \ti^5 \mphi, 
\end{align}
where we use $\qel \sim \alpha^2 T_s^3$. 
The thermal potential is now determined by the temperature of the soft sector: 
\begin{align}
	 k_\text{max} \sim T_* \sim T_s \sim \alpha^4 \Gphi \ti m_I.
\end{align}

Note that the distribution function for soft $\chi$-particles after $\ti > \ti_\text{soft}$ is given by
\begin{align}
f_s (t, k) \sim
\begin{cases}
	\cfrac{T_s}{k} 
	& \text{for}~~ k \lesssim T_s, \\[1em]
	\left( \cfrac{\ti_\text{soft}}{\ti} \right)^2 \left( \cfrac{T_s}{k} \right)^{7/2}
	&\text{for}~~ T_s \lesssim k \lesssim k_\text{form}, \\[1em]
	\alpha \tilde \Gamma_I \tilde t^{-1} \left( \cfrac{m_I}{k} \right)^{3}
	&\text{for}~~ k_\text{form} \lesssim k.
\end{cases}
\label{f_s after t_soft}
\end{align}
Also note that the ``hard'' interactions among soft particles with the momentum exchange of $T_s$
is faster than the cosmic expansion:
\begin{align}
	\frac{\alpha^2 n_s / T_s^2}{H} \sim \left( \frac{\ti}{\ti_\text{soft}} \right)^2 > 1.
	\label{eq:softvsh}
\end{align}
Hence, the soft sector is thermalized separately.

\begin{figure}[t]
\centering 
\includegraphics[width=.7\textwidth
]{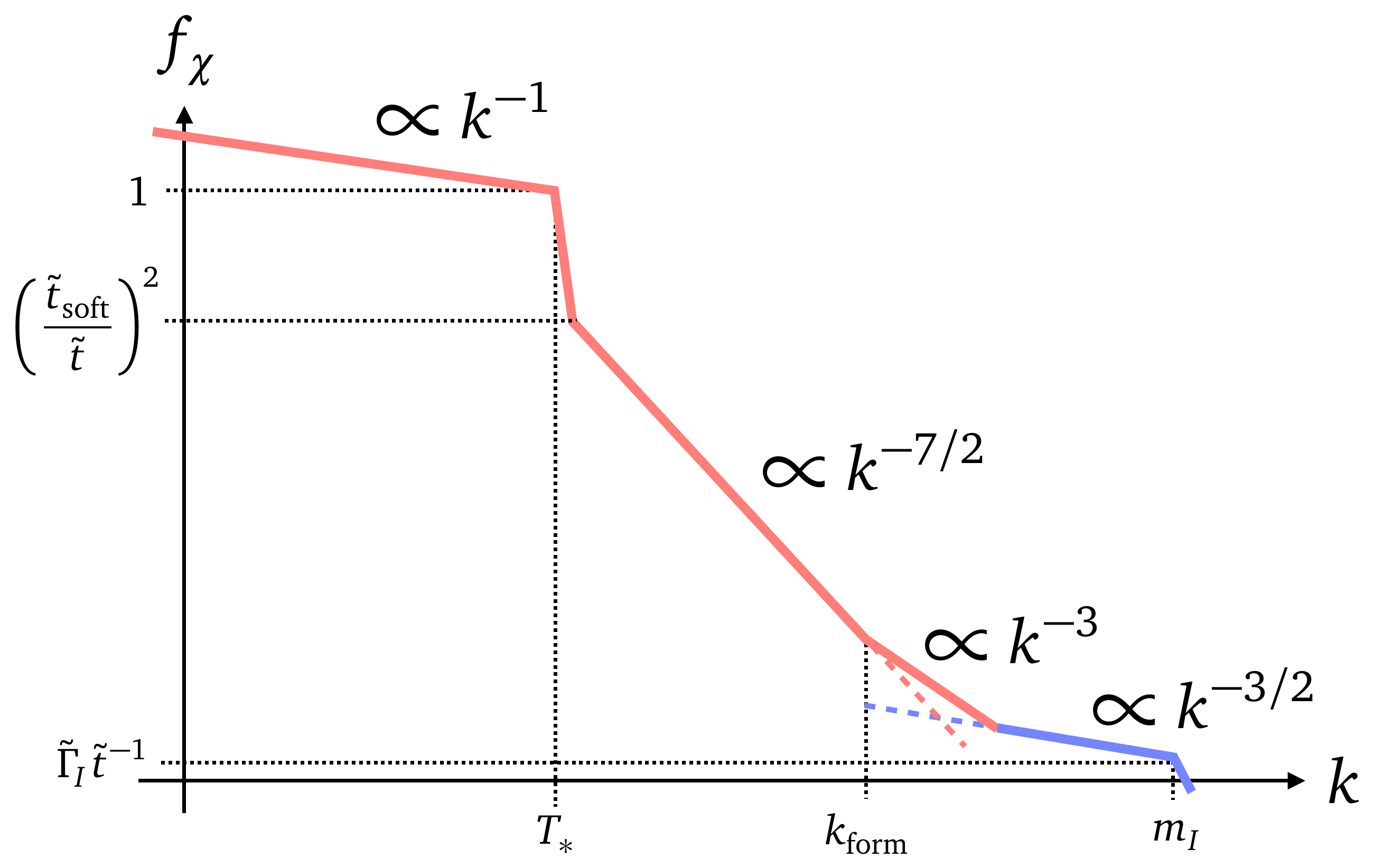} 
\caption{\small
Distribution of $\chi$-particles for $\ti_{\rm soft} < \ti < \ti_{\rm max}$. 
Recall that $ T_\ast \sim \alpha \tilde \Gamma_I^{1/2} m_I (\tilde t/ \tilde t_\text{soft})$,
$k_\text{form} \sim \alpha^{-2} T_\ast (\tilde t / \tilde t_\text{soft})^4$,
$t_\text{soft} \equiv \alpha^{-3} \tilde \Gamma_I^{-1/2}$,
and $t_\text{max} \equiv \alpha^{-16/5} \Gphi^{-3/5}$ [See Eq.~\eqref{eq:max}].
{\bf Soft} ({\color{ppink}pink}):
The thermal-like distribution dominates below $T_\ast (= T_s)$,
the LPM-suppressed spectrum can be seen between $k_\text{max}$ and $ k_\text{form}$,
and the vacuum cascade may be relevant above $k_\text{form}$.
{\bf Hard} ({\color{bblue}blue}):
The hard spectrum is sourced by the direct decay of inflaton and
by its red-shifted spectrum of previously produced one.
Eventually, 
the distributions of soft and hard sectors become to coincide with each other 
at the wavenumber of $k \sim m_I (\ti / \ti_{\rm max})^{5/4}$.
}
  \label{distribution2}
\end{figure}

We illustrate the distribution of $\chi$-particles in Fig.~\ref{distribution2}, 
where we include the hard sector as well as the soft one. 
The distribution of the soft sector is given by Eq.~(\ref{f_s after t_soft}), 
while that of the hard sector is given by Eq.~(\ref{hard-distribution}). 
Eventually, 
the distributions of soft and hard sectors coincide with each other 
at the wavenumber of $k \sim m_I (\ti / \ti_{\rm max})^{5/4}$, 
where $\ti_{\rm max} \equiv \alpha^{-16/5} \Gphi^{-3/5}$ is defined in the next subsection
[See Eq.~\eqref{eq:max}].

\subsubsection*{Production of $\chi_\phi$-particles}

If the effective mass term, $|g_\phi \phi |$,  is smaller than the temperature of the soft sector, $|g_\phi \phi| < T_s$,
then the dominant contribution to the distribution function of $\chi_\phi$-particles is given by
\begin{align}
	f_{\chi_\phi} (t, k) \sim \frac{T_s}{k}~~ \text{for}~~ k < T_s,
	\label{f_chi_phi1}
\end{align}
which yields
\begin{align}
	\left. T_{\ast, \chi_\phi}^2 \right|_\text{indir} \sim T_s^2,
	\label{T_*_chi_phi1}
\end{align}
for $|g_\phi \phi| < T_s$.
Basically, this is because ``hard'' interactions among the soft sector with the momentum exchange
of $T_s$ is much faster than the cosmic expansion at this stage,
as shown in Eq.~\eqref{eq:softvsh}.
Thus, $\chi_\phi$-particles with a mass lighter than $T_s$ are efficiently produced
and participate in the thermal plasma.

The case of $|g_\phi \phi| > T_s$ is calculated in Appendix. 
The $\chi_\phi$-particle production processes are the same with the ones in the previous subsection, 
though the number densities of $n_s$ and $n_h$ are different.

\subsubsection*{Thermal potential}
Here we summarize the thermal potentials 
and discuss which contribution dominates for each case. 
See also Appendix~\ref{app:eff_t}.
Again, note that for an interval between $T_\ast < | g_\phi \phi | < \alpha^{-1} T_\ast$,
the effective temperature may depend on $\epsilon^2$.
We do not care about this small interval for our rough estimation.

\begin{itemize}
\item{Indirect $\chi_\phi$-production (produced from hard primaries/soft daughters):}
\begin{itemize}
\item
First, we consider a large field value regime, $| g_\phi \phi | > \alpha^{-1} T_\ast$. 
In the case of $\Gphi \lesssim \alpha^6$, 
the effective mass from abundant $\chi_\phi$-particles always gives subdominant corrections to 
the effective potential, meanwhile the log contribution dominates:
\begin{align}
	\alpha^2 \frac{T_\ast^4}{\phi^2} \sim
	\alpha_\phi \alpha^2 T_\ast^2 \left( \frac{T_\ast}{|g_\phi \phi|} \right)^{2}.
	\label{eq:log_p1_idir_l2}
\end{align}
Even in the case of $\Gphi \gtrsim \alpha^6$, 
the effect of $\chi_\phi$-particles is very limited, 
so that we can neglect its contribution for a rough estimation. 
For $\ti \gtrsim \Min\, [ \alpha^{-1} \ti_{\rm soft}, t_\text{max} ]$,
that is, soon after the soft sector is thermalized
the running coupling constant contribution dominates for $|g_\phi \phi| > \alpha^{-1} T_\ast$ 
independently of $\Gphi$. 
This is because the number density of soft sector increases with time,
while the number density of hard primaries decreases due to the red-shift of inflaton number density.

\item
For a small field value regime, $| g_\phi \phi | < T_\ast$,
the effective mass from abundant $\chi_\phi$-particles dominates the effective potential:
\begin{align}
	\alpha_\phi \left. T_{\ast, \chi_\phi}^2 \right|_\text{indir} = \alpha_\phi T_\ast^2.
\end{align}
\end{itemize}
\item{Direct $\chi_\phi$-production (produced from inflaton decay):}
\begin{itemize}
\item
For the large field value regime, $| g_\phi \phi | > \alpha^{-1} T_\ast$, 
the effective potential is given by the competition of two contributions;
the log dependence of the running coupling constant and hard $\chi_\phi$-primaries from the direct decay of inflaton.
The contribution from the direct decay of inflaton, which is given by
\begin{align}
	\left. T_{\ast, \chi_\phi}^2 \right|_\text{dir} &\sim
	\alpha \Gphi^{1/2} T_\ast^2  \left( \frac{\ti_\text{soft}}{\ti} \right)^{3}\,
	\Min \left[ 
		1, \left( \frac{\alpha}{\epsilon^2} \right) \left( \frac{T_\ast}{g_\phi \phi} \right) \left( \frac{\ti_\text{soft}}{\ti} \right)^2
	\right],
	\label{eq:p2_dir}
\end{align}
dominates the effective temperature for
\begin{align}
	m_I > |g_\phi \phi| > \alpha m_I \left( \frac{\ti}{\ti_\text{soft}} \right)^{5/2} 
	\Min \left[ \alpha^{1/2} \Gphi^{1/4}, \alpha\left( \frac{\epsilon^2}{\alpha} \right) \left( \frac{\ti}{\ti_\text{soft}} \right)^{7/2}  \right].
\end{align}
\item
For the small field value regime, $| g_\phi \phi | <  T_\ast$, 
the contribution from abundant $\chi_\phi$-particles is always larger than other ones;
from the running coupling constant and also form the direct decay of inflaton:
\begin{align}
	\alpha_\phi  T_{\ast, \chi_\phi}^2  \simeq  \alpha_\phi \left. T_{\ast, \chi_\phi}^2  \right|_\text{indir}
	=  \alpha_\phi T_{\ast, \chi}^2.
\end{align}
This is simply because the effective temperature for massless modes is dominated by the soft sector
as shown in Eq.~\eqref{eq:effT_softvshard}.
\end{itemize}
\end{itemize}

\subsection{$\ti_{\rm max} < \ti < \ti_{\rm RH}$}

When $\Gamma_{\rm split} (k \simeq m_\phi) \sim H$ is satisfied, 
that is, when $k_{\rm split}$ becomes comparable to $\mphi$, 
primary particles lose their energy completely and are thermalized soon. 
This occurs at the time of 
\begin{align}
	 \ti \sim \ti_{\rm max} \equiv \alpha^{-16/5} \Gphi^{-3/5}. 
	 \label{eq:max}
\end{align}
This is the actual thermalization time scale of hard particles.

\subsubsection*{Production of $\chi$-particles}

The energy conservation implies that 
\begin{align}
	 \rho_r \sim \rho_\phi \Gamma_\phi t \sim H(t) \Gamma_\phi \Mpl^2, 
\end{align}
where $\rho_r$ ($\sim T_\ast^4$) is the energy density of radiation. 
This gives us the time evolution of the temperature such as 
\begin{align} 
	 T_\ast \sim \Gphi^{1/4} \ti^{-1/4} \mphi. 
	 \label{T before reheating}
\end{align}
Since the hard sector as well as the soft sector are thermalized, 
they can be described by a single parameter $T_\ast$.

\begin{figure}[t]
\centering 
\includegraphics[width=.8\textwidth
]{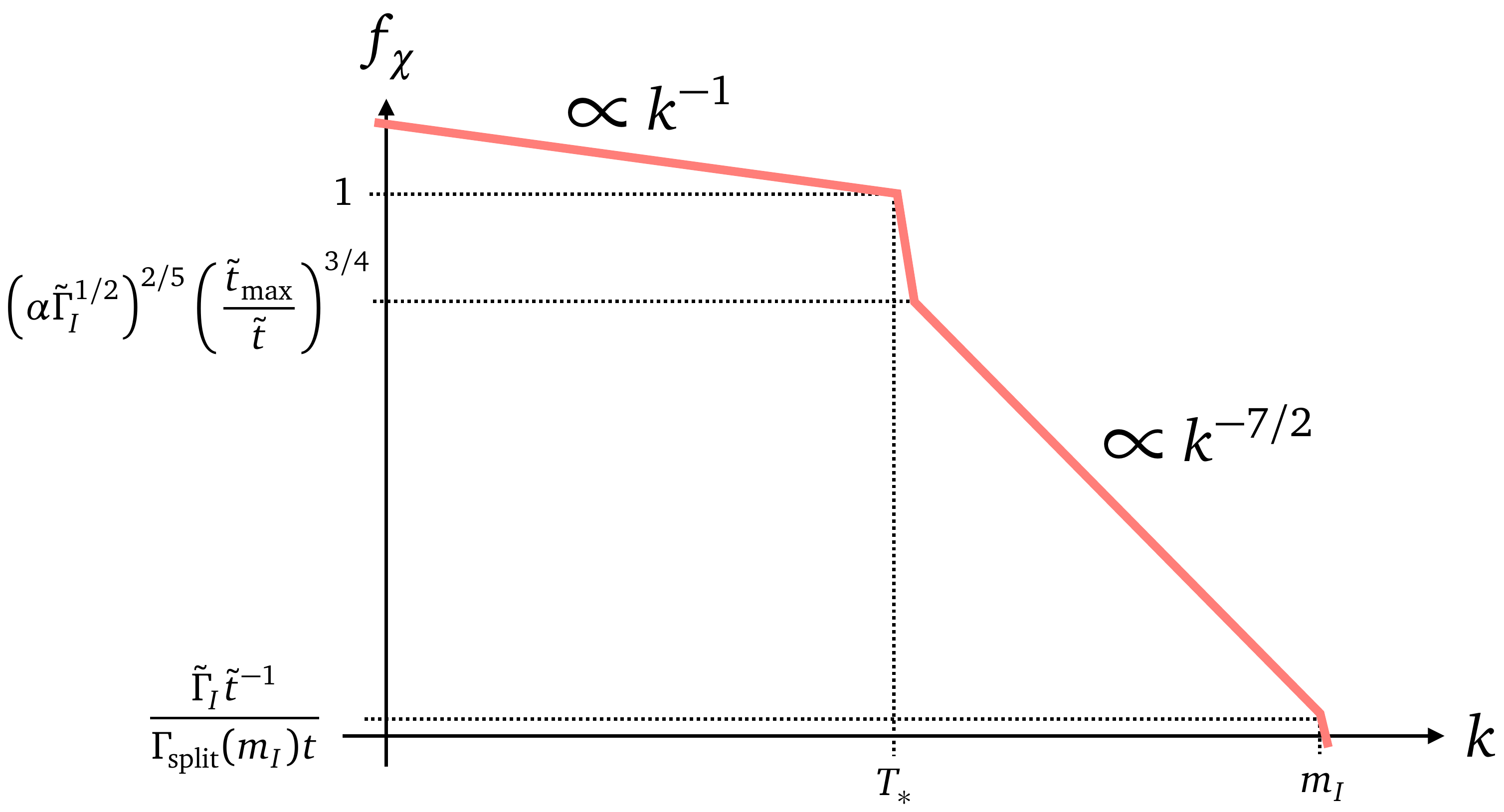} 
\caption{\small
Distribution of $\chi$-particles for $\ti_{\rm max} < \ti < \ti_{\rm RH}$. 
Recall that $ T_\ast \sim \Gphi^{1/4} \tilde t^{-1/4} m_I$,
$t_\text{max} = \equiv \alpha^{-16/5} \Gphi^{-3/5}$,
and $t_\text{RH} = \Gphi^{-1} \Mpl^2/m_I^2$
[See Eq.~\eqref{eq:RH}].
Note that there remains the tail of LPM-suppressed spectrum $\propto k^{-7/2}$
because the inflaton continuously produces primaries with the momentum of $p \sim m_I$.
}
  \label{distribution3}
\end{figure}

We illustrate the distribution of $\chi$-particles in Fig.~\ref{distribution3}. 
The distribution of smaller wavenumber modes than $T_\ast$ 
is a thermal one, 
while that of larger wavenumber modes 
is determined by the LPM effect as Eq.~(\ref{f_s}). 
Note that the hard sector is completely thermalized, 
so that it is absent in Fig.~\ref{distribution3}.

\subsubsection*{Production of $\chi_\phi$-particles and thermal potential}

The production process of $\chi_\phi$-particles 
is the same with the one considered in the previous subsection.
However, since the number density of hard primaries 
decreases efficiently by splittings for $t > t_{\rm max}$, 
the contribution from the direct decay of inflaton
is now given by
\begin{align}
	\left. f_{\chi_\phi} (t, m_I) \right|_\text{dir} \sim \Gphi \ti^{-1} \,
	\Min \left[ \frac{1}{\Gamma_\text{split} (m_I) t}, \frac{1}{\Gamma_\text{decay} t} \right].
\end{align}

Next, we discuss which contribution dominates the thermal potential. 
See also Appendix~\ref{app:eff_t}. 
Again, note that for an interval between $T_s < | g_\phi \phi |  < \alpha^{-1} T_s$,
the effective temperature may depend on $\epsilon^2$.
We do not care about this small interval for our rough estimation.

\begin{itemize}
\item{Indirect $\chi_\phi$-production (produced from hard primaries/soft daughters):}
\begin{itemize}
\item For the large field value regime, $|g_\phi \phi| > \alpha^{-1} T_s$,
we find that 
the effect on the effective potential from $\chi_\phi$-particles 
is always subdominant compared with the thermal log potential of Eq.~(\ref{eq:log}). 
\item For the small field value regime, $|g_\phi \phi| < T_s$,
the distribution of $\chi_\phi$-particles and its effective temperature are given by Eqs.~(\ref{f_chi_phi1}) and (\ref{T_*_chi_phi1}). 
In this case, the effective potential of $\phi$ is given by the usual thermal mass term of Eq.~(\ref{eq:mass_2}).
\end{itemize}
\item{Direct $\chi_\phi$-production (produced from inflaton decay):}
\begin{itemize}
\item For the large field value regime, $|g_\phi \phi| > \alpha^{-1} T_s$,
the effective potential is determined  by the competition between the contribution from
the running coupling constant and the direct decay of inflaton.
The latter yields
\begin{align}
	\left. T_{\ast, \chi_\phi}^2 \right|_\text{dir} 
	\sim \left( \alpha \Gphi^{1/2} \right)^{8/5} \left( \frac{\ti_\text{max}}{\ti} \right)^{9/8} T_\ast^2 \,
	\Min \left[
		1,  \left( \alpha \Gphi^{1/2} \right)^{2/5} \left( \frac{\alpha}{\epsilon^2} \right)
		\left( \frac{T_\ast}{g_\phi \phi} \right) \left( \frac{\ti_\text{max}}{\ti} \right)^{1/8}
	\right].
	\label{eq:p3_dir}
\end{align}
This contribution dominates the effective potential for
\begin{align}
	m_I > | g_\phi \phi | >
	\alpha m_I \left( \frac{\ti}{\ti_\text{max}} \right)^{5/16}
	\Max \left[
		1, \left( \alpha \Gphi^{1/2} \right)^{-6/5} \epsilon^2 \left( \frac{\ti}{\ti_\text{max}} \right)^{11/16}
	\right].
\end{align}
\item For the small field value regime, $|g_\phi \phi| < T_s$,
the effective potential of $\phi$ is governed by the usual thermal mass term of Eq.~(\ref{eq:mass_2}).
\end{itemize}
\end{itemize}

\subsection{$\ti_{\rm RH} < \ti$}

When $\Gamma_I \sim H(t)$ is satisfied, 
the energy density of radiation becomes to dominate that of the Universe. 
This is when the reheating is completed:
\begin{align}
	 \ti \sim \ti_{\rm RH} \equiv \Gphi^{-1} \frac{\Mpl^2}{m_I^2}. 
	 \label{eq:RH}
\end{align}
The reheating temperature is given by 
\begin{align}
	 T_{\rm RH} \sim \sqrt{ \Gamma_I \Mpl}. 
\end{align} 
After the reheating is completed, 
the temperature of the Universe decreases with time as 
\begin{align}
	 T_\ast \sim 
	 T_{\rm RH} 
	 \lmk \frac{\ti_{\rm RH}}{\ti} \rmk^{1/2}, 
\end{align}
because $H^2 \sim T_\ast^4 / \Mpl^2$ in the radiation dominated era.

\subsection{Summary \label{summary in sec 3}}

\begin{figure}[t]
\centering 
\includegraphics[width=.7\textwidth
]{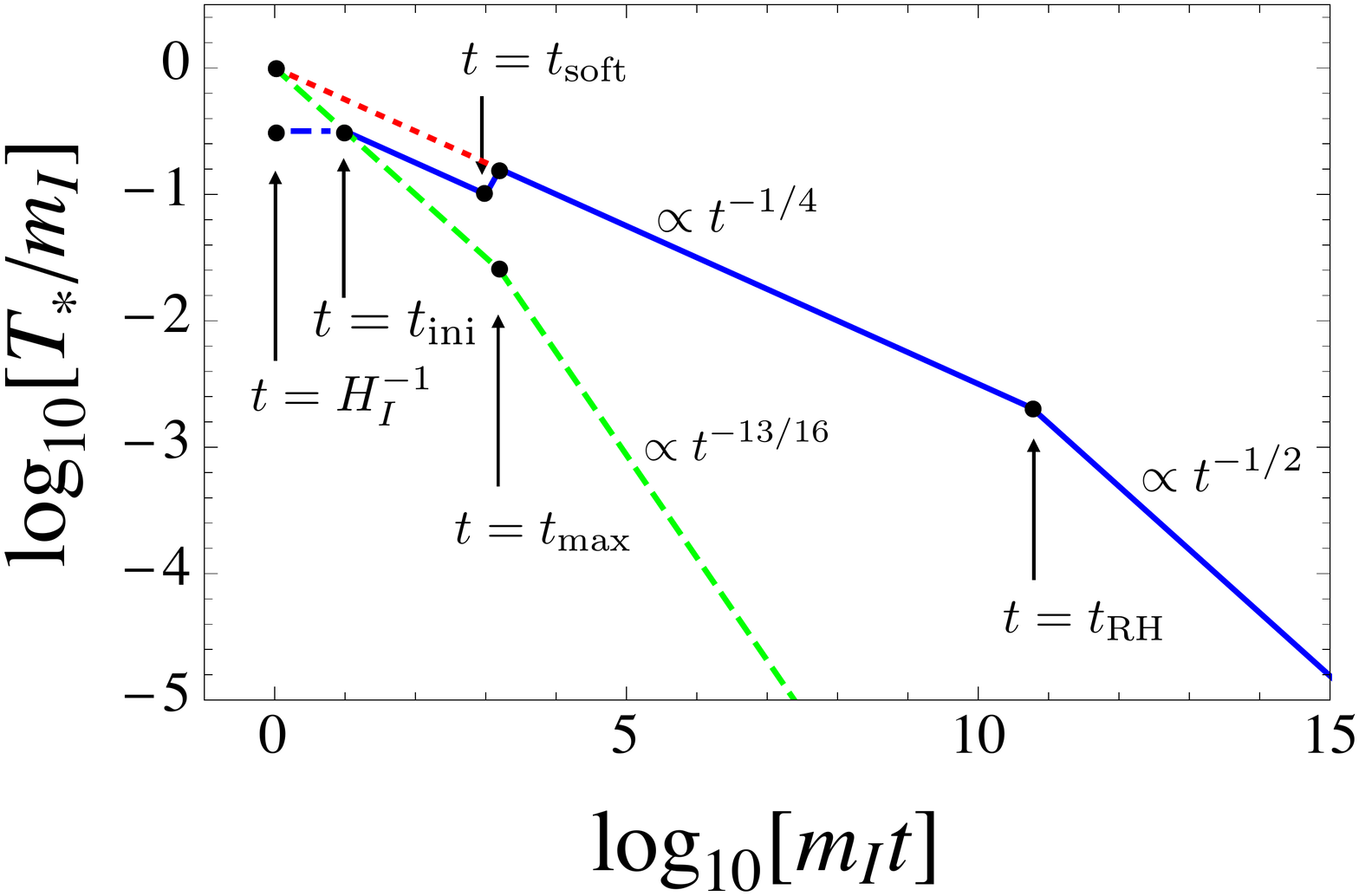}
\caption{\small
Evolution plots for the effective temperature of the soft sector $T_\ast$. 
We assume $\alpha = 0.1$, $H_I = m_I$, and $m_I = 10^{13} \GEV$. 
We take $\Gphi = 1$, 
which corresponds to the case of $\Gphi > \alpha^3$. 
The blue lines are results derived in this paper, 
while the red dotted lines describe the temperature derived in the literature by assuming 
the ``{\it instantaneous thermalization}''. 
We also show the effective temperature of the hard sector as green dashed lines, 
for the case that they are produced from inflaton decay,
where we neglect the decay of $\chi_\phi$-particles. 
The blue dot-dashed line is the effective temperature for $\ti < \ti_{\rm ini}$, 
which is estimated from the discussion of the vacuum cascades [see Eq.~(\ref{f_s before t_ini})]. 
}
  \label{fig1}
\end{figure}

\begin{figure}[t]
\centering 
\includegraphics[width=.7\textwidth
]{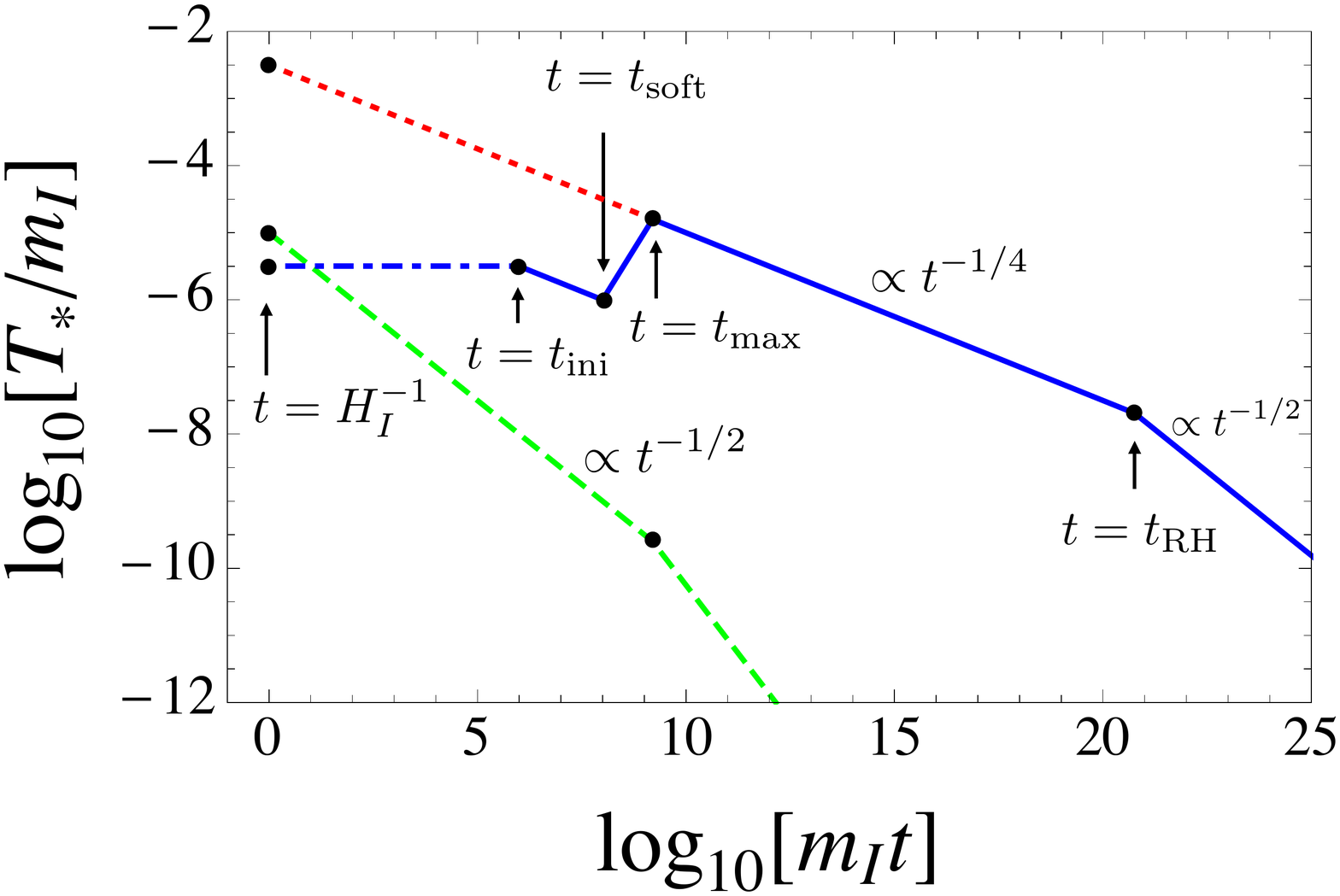} 
\caption{\small
Same as Fig.~\ref{fig1}, but 
we take
$\Gphi = 10^{-10}$, 
which corresponds to the case of $\Gphi < \alpha^3$. 
}
  \label{fig2}
\end{figure}

To sum up, 
we obtain the evolution of the effective temperature of
$\chi$-particles
as follows: 
\begin{align}
	\frac{T_*}{\mphi} \sim 
	\begin{cases}
		\alpha^{1/2} \Gphi^{1/2} \lmk \cfrac{\ti}{\ti_{\rm ini}} \rmk^{-1/4} &\text{for}~~  
		\ti_{\rm ini} \lesssim \ti \lesssim \ti_{\rm soft} \\[1em]
		\alpha \Gphi^{1/2} \lmk \cfrac{\ti}{\ti_{\rm soft}} \rmk 
		&\text{for}~~  \ti_{\rm soft} \lesssim \ti \lesssim \ti_{\rm max} \\[1em]
		\alpha^{4/5} \Gphi^{2/5} \lmk \cfrac{\ti}{\ti_{\rm max}} \rmk^{-1/4} 
		&\text{for}~~  \ti_{\rm max}  \lesssim \ti \lesssim \ti_{\rm RH} 
	\end{cases}
\end{align}
where the time scales are given by 
\begin{align}
	 \ti_{\rm ini} &\equiv \alpha^{-1} \Gphi^{-1/2}, \\
	 \ti_{\rm soft} &\equiv \alpha^{-3} \Gphi^{-1/2}, \\
	 \ti_{\rm max} &\equiv \alpha^{-16/5} \Gphi^{-3/5}, \\
	 \ti_{\rm RH} &\equiv \Gphi^{-1} \frac{\Mpl^2}{m_I^2}. 
\end{align}
Here note that we have implicitly assumed $\ti_\text{RH} > \ti_\text{max}$ so far,
which implies the thermalization takes place faster than the complete decay of inflaton.
As shown in Ref.~\cite{Harigaya:2013vwa},
this inequality is satisfied in most cases: $\alpha \gtrsim 4 \times 10^{-4} (m_I / 10^{13}\GEV)^{5/8} \Gphi^{1/8}$.
These results are illustrated in Figs.~\ref{fig1} and \ref{fig2} as blue lines. 
The effective temperature of the soft sector has two local maxima 
as $T_*/\mphi \sim \alpha^{1/2} \Gphi^{1/2}$ and $\alpha^{4/5} \Gphi^{2/5}$ 
at the time of 
$\ti = \ti_{\rm ini}$ and $\ti_{\rm max}$, respectively. 
The global maximum of the effective temperature is given by 
\begin{align}
	\left. T_* \right\vert_{\rm max} \sim 
	\begin{cases}
		\alpha^{1/2} \Gphi^{1/2} \mphi 
		&\text{for}~~ 
		1 \gtrsim \Gphi \gtrsim \alpha^3 \\[.5em]
		\alpha^{4/5} \Gphi^{2/5} \mphi
		&\text{for}~~ 
		\Gphi \lesssim \alpha^3 
	\end{cases}
	\label{max temp}
\end{align}
There is a local minimum given as $T_* \sim \alpha \Gphi^{1/2}$ 
at $\ti \sim \ti_{\rm soft}$. 
Note that the effective temperature for $\ti < \ti_{\rm ini}$ is at most that for $\ti \sim \ti_{\rm ini}$ 
from the discussion of the vacuum cascades [see Eq.~(\ref{f_s before t_ini})], 
so that the maximal temperature is in fact given by Eq.~(\ref{max temp}). 
The thermalization process described in this section 
is summarized in the second paragraph in Sec.~\ref{sec1-2}. 

Now we obtain the evolution of the thermal effects on the potential of $\phi$. 
If $\chi_\phi$-particles are not directly produced from inflaton decay, 
the thermal potential of $\phi$ is roughly given by 
\begin{align}
	 V_{\rm eff} (\phi) \sim 
	\begin{cases}
		 \alpha_\phi T_*^2 \phi^2 
		 &\text{for}~~ 
		 | g_ \phi \phi | \ll T_* \\[.5em]
		 \alpha^2 T_*^4 \log \lmk \cfrac{\phi^2}{T_*^2} \rmk 
		 &\text{for}~~ 
		 |g_\phi \phi | \gg T_*, 
	\end{cases}
\end{align}
for the case of $\Gphi \lesssim \alpha^6$. 
Here, we neglect the complicated parameter dependence of the effective temperature $T_{\ast, \chi_\phi}$
for $\alpha T_\ast \lesssim | g_\phi \phi | \lesssim \alpha^{-1} T_\ast$,
and roughly evaluate it as $T_{\ast, \chi_\phi} \sim T_\ast$ for $|g_\phi \phi| \ll T_\ast$.
Even in the case of $\Gphi \gtrsim \alpha^6$, 
the deviation of the above formula is very limited, 
so that one can still use it for a rough estimation. 
On the other hand, 
if $\chi_\phi$-particles are directly produced from inflaton decay, 
it may affect the thermal potential of $\phi$. 
In particular, 
for the large field value regime, $| g_\phi \phi | \gtrsim \alpha^{-1} T_*$, 
the effective potential is given by the competition of two contributions; 
the thermal log potential from $\chi$-particles and thermal mass from $\chi_\phi$-particles. 
The latter contribution is given in Eqs.~\eqref{eq:p1_dir}, \eqref{eq:p2_dir} and \eqref{eq:p3_dir}. 
If we can neglect the decay of $\chi_\phi$-particles owing to the smallness of $\epsilon$, 
$T_{\ast, \chi_\phi}$ evolves as green dashed lines in Figs.~\ref{fig1} and \ref{fig2}.

Finally, we comment on the uncertainty of our results. 
We omit the log-enhancement factor of splitting rate in Eq.~(\ref{splitting rate}) for simplicity, which may be as large as a factor of ten,
and also a model dependent factor.
See App.~\ref{app:lpm} for this issue.
There is also numerical factor that is derived from numerical calculations of Boltzmann equations in Ref.~\cite{Kurkela:2014tea}.
Here we collectively denote these factor as $c$, i.e., we write $\Gamma_{\rm split} (k) = c \alpha \sqrt{\hat{q}_{\rm el} / k}$ 
and discuss the uncertainty of the maximal temperature of the Universe. 
Since we are mostly interested in the case of $\Gphi \ll \alpha^3 \ll 1$, 
we focus on the effective termperature at the time of $\ti \sim \ti_{\rm max}$. 
The effective temperature $T_*$ becomes the maximal value 
at the time when $\Gamma_{\rm spilt} (k \simeq m_I) \simeq H$ is satisfied. 
Note here that there are another two model dependent factors:
(i) total degrees of freedoms of plasma, $g_\ast$, and
(ii) a factor of the diffusion coefficient, $\hat q_\text{el} = c_q \alpha^2 T^3$ [See Eq.~\eqref{eq:diff_coeff}].
As a result,
the maximal temperature is given by the second line of Eq.~(\ref{max temp}) 
times the uncertainty factor of $[( 36 / \pi^2 g_\ast )c c_q^{1/2}]^{2/5}$. 
This dependence is rather small 
because the time-dependence of the effective temperature has a small power of $-1/4$. 
We may conclude that 
the uncertainty of the maximal temperature is at most a factor of ten.

\section{\label{applications}Applications}

In this section,
we discuss implications of obtained results to the dynamics of scalar condensates
in the early Universe.
In particular, we focus on their effects on SSB and onset of oscillation,
which are important to determine the fate of the Universe.

Suppose that $\phi$ has a tachyonic mass at zero-temperature 
and is a field responsible for the SSB of some symmetry. 
When it acquires a thermal mass larger than its zero-temperature tachyonic mass, 
it is stabilized at the origin of the potential. 
As the temperature decreases, 
the thermal mass becomes comparable to its zero-temperature mass. 
Then $\phi$ starts to oscillate around the true vacuum and obtains a nonzero VEV, 
so that the SSB occurs at that time. 
The SSB may results in formation of topological defects, such as cosmic strings 
and domain walls. 
These topological defects 
may predict some detectable signals. 
In particular, 
the energy density of domain walls decreases as $a^{-2}$, 
so that they eventually dominate the Universe and spoil the success of the standard cosmology~\cite{Zeldovich:1974uw}. 
However, 
when the SSB occurs before inflation, 
these topological defects are washed out by inflation. 
Thus, 
it is important to determine the condition that the SSB occurs after inflation. 

Also, a scalar field can have a large expectation value during inflation
due to the negative Hubble induced mass term~\cite{Dine:1995kz}.
Such a scalar condensate starts to oscillate around its effective potential minimum
when the potential force becomes comparable to the Hubble friction.
The subsequent evolution of the Universe can crucially depend on the time
when the scalar field starts to oscillate.
For instance, if its oscillation starts earlier due to the finite density effects, then the scalar field tends to dissipate its 
energy earlier into the background plasma.
More importantly,
the baryon asymmetry of the Universe is fixed at the time when
the AD field starts to oscillate, as we will see later.

In the next subsection, 
we calculate the condition that the SSB occurs after inflation. 
We apply the result to a QCD axion model, which is well motivated in light of solution to the strong CP problem. 
Then we also apply the result to the electroweak symmetry breaking 
and comment on the DM production from the thermal plasma. 
Finally, we consider the dynamics of a flat direction in SUSY theories, 
especially in the context of the Affleck-Dine baryogenesis.

\subsection{\label{PQ symmetry}PQ symmetry}

In this section, we consider the following potential for the field $\phi$: 
\begin{align}
	 V(\phi) = m_T^2 \abs{\phi}^2 + \frac{\lambda^2}{2} \lmk \abs{\phi}^2- v^2 \rmk^2, 
\end{align}
where we include the thermal mass term ($m_T^2 \sim \alpha_\phi T_{\ast, \chi_\phi}^2$). 
The SSB occurs at the effective temperature of $T_{\rm SSB}$ that is determined by 
$m_T (T_{\rm SSB}) \simeq \lambda v$.

When we require $m_T (T) < \lambda v$ throughout the history of the Universe, 
we obtain an upper bound on the reheating temperature as 
\begin{align}
	 T_{\rm RH} \lesssim 
	 2 \times 10^{10} \GEV 
	 \lmk \frac{\alpha}{0.1} \rmk^{-1} 
	 \lmk \frac{\lambda v}{10^{12} \GEV} \rmk 
	 \lmk \frac{m_\phi}{10^{13} \GEV} \rmk^{1/2}, 
	 \label{T_RH upper bound} 
\end{align}
where we assume that $\Gphi > \alpha^3$,
$\alpha \sim \alpha_\phi$, and that particles which couple to $\phi$ (PQ-quarks)
are not produced  directly from inflaton decay.
This implies that the reheating temperature has to be much lower than the dynamical 
scale $v$ to avoid the symmetry restoration of that symmetry.\footnote{
	This is merely a necessary condition to avoid the symmetry restoration.
	If the field value during inflation is much larger than $v$ due to the negative Hubble induced mass,
	the non-thermal phase transition can occur,
	which may result in formation of topological defects~\cite{Kofman:1995fi,Khlebnikov:1998sz,Tkachev:1995md,Felder:2000sf}.
	To study its fate, one may also have to take into account of background plasma,
	which is involved and beyond the scope of this paper.
	See \cite{Moroi:2013tea} for instance.
}

Here, let us consider a QCD axion model with right-handed neutrinos~\cite{Yanagida:1979as,Yanagida:1980xy,GellMann:1980vs,Minkowski:1977sc} 
and identify $\phi$ as the field responsible to the SSB of PQ symmetry. 
When the SSB occurs after inflation, 
cosmic strings form at the phase transition~\cite{Vilenkin:1982ks}. 
After the QCD phase transition, 
the non-perturbative effect associated with instantons 
breaks $U(1)_{\rm PQ}$ symmetry down to $Z_n$, 
where $n$ is an integer depending on models. 
This implies that domain walls form at the QCD phase transition. 
While these domain walls are short lived in the case of $n=1$~\cite{Kim:1979if, Shifman:1979if, Lazarides:1982tw, Kawasaki:2015ofa}, 
they are stable and disastrous in the case of $n \ge 2$~\cite{Zeldovich:1974uw, Sikivie:1982qv}. 
One of the simplest solution of this domain wall problem 
is that the PQ symmetry is never restored after inflation.\footnote{
	In this case, 
	the axion DM may predict sizable isocurvature fluctuations~\cite{Axenides:1983hj, Seckel:1985tj, Turner:1990uz}, 
	which is tightly constrained by the observations of CMB fluctuations. 
	This severely restricts the energy scale of inflation 
	to suppress isocurvature fluctuations. 
}
This scenario requires a reheating temperature lower than the one derived in Eq.~(\ref{T_RH upper bound}). 
QCD axion models predict a pseudo-NG boson called axion, 
which is a good candidate of DM. 
The abundance of axion is related to the PQ breaking scale $v$, 
so that the observed DM abundance determines its value~\cite{Preskill:1982cy, Abbott:1982af, Dine:1982ah}. 
The result is given as 
\begin{align}
	 v \simeq 8 \times 10^{11} \GEV  \times n \abs{\theta_0}^2, 
\end{align}
where $\theta_0$ is the initial phase of axion field and $n$ is the domain wall number. 
This implies that the reheating temperature should be lower than $10^{10} \GEV$ 
[see Eq.~\eqref{T_RH upper bound}]. 
Hence,
leptogenesis may be marginally realized to explain the baryon asymmetry of the Universe~\cite{Fukugita:1986hr} (see also Ref.~\cite{Buchmuller:2005eh}).

\subsection{\label{EWPT}Electroweak phase transition 
and DM production
}

In order not to spoil the success of the BBN, 
reheating has to be completed before the BBN epoch, 
which means $T_{\rm RH} \gtrsim 1 \MEV$. 
This requires that $\Gphi$ has to be larger than the following value: 
\begin{align}
	 \Gphi \sim \frac{T_{\rm RH}^2 \Mpl}{m_I^3} 
	 \sim 10^{-27} 
	 \lmk \frac{T_{\rm RH}}{1 \MEV} \rmk^2 
	 \lmk \frac{m_I}{10^{13} \GEV} \rmk^{-3}. 
\end{align}
In this case, 
the effective temperature of the soft sector 
can be 
as large as 
\begin{align}
	 \left. T_* \right\vert_{\rm max}
	 &\sim  \alpha^{4/5} \Gphi^{2/5} m_I \\ 
	 &\sim 
	 200 \GEV \times 
	  \alpha^{4/5} 
	 \lmk \frac{\Gphi}{10^{-27}} \rmk^{2/5} 
	 \lmk \frac{m_I}{10^{13} \GEV} \rmk. 
	 \label{m_T for Higgs}
\end{align}
Assuming 
$\alpha = 10^{-1}$, 
we obtain 
$\left. T_* \right\vert_{\rm max} \sim 30 \GEV$ for the case of $m_I = 10^{13} \GEV$. 
Even if $\chi_\phi$-particles are generated directly from the inflaton decay, 
the effective temperature of the hard sector is at most 
\begin{align}
	 \left. T_{*, \chi_\phi} \right\vert_{\text{indir}} 
	 &\sim \Gphi^{1/2} m_I \\
	 &\sim 0.3 \GEV \times \lmk \frac{\Gphi}{10^{-27}} \rmk^{1/2} 
	  \lmk \frac{m_I}{10^{13} \GEV} \rmk, 
	\label{m_T^hard}
\end{align}
just after inflation, where we assume $\ti \sim 1$.

Here, let us consider thermal production of DM in a low reheating temperature. 
Even if the reheating temperature is lower than the mass of DM, 
it can be produced from the thermal plasma before the reheating is completed~\cite{Chung:1998rq, Giudice:2000ex}. 
The present energy density of DM from the thermal production
divided by the entropy density of the Universe is calculated as 
\begin{align}
	 \left. \frac{\rho_{\rm DM}^{\rm th}}{s} \right\vert_{\text{now}}
	 \simeq \left. \frac{\rho_{\rm DM}^{\rm th}}{s} \right\vert_{\text{F}} \lmk \frac{T_{\rm RH}}{T_{\text{F}}} \rmk^{5},
	\label{thermal production of DM}
\end{align}
where the subscript ``F'' represents the corresponding value at the time of DM freeze-out 
(see, {\it e.g.,} Ref.~\cite{Harigaya:2014waa}). 
To derive this, 
it is assumed that the maximal temperature of the Universe after inflation 
is larger than the mass of DM. 
However, Eqs.~(\ref{m_T for Higgs}) and (\ref{m_T^hard}) 
show that 
the maximal temperature of the Universe is at most the electroweak scale 
for $T_{\rm RH} = {\cal O}(1) \MEV$. 
This means that in such a low reheating temperature 
the thermal production of DM is not efficient 
and Eq.~(\ref{thermal production of DM}) cannot be applicable to calculate the amount of DM. 
Even in this case, 
DM is produced from interactions between the hard and soft sector 
and this contribution is actually much more efficient to generate DM~\cite{Harigaya:2014waa}.

Next, let us identify the field $\phi$ as the Higgs boson~\cite{Kirzhnits:1972iw, Kirzhnits:1972ut, Dolan:1973qd, Weinberg:1974hy, Kirzhnits:1976ts}. 
Then $\chi_\phi$-particles are identified as the quarks, leptons, and $SU(2)_L \times U(1)_Y$ gauge bosons. 
In general, 
they are generated directly from the inflaton decay, 
so that 
the case of ``direct $\chi_\phi$-production'' should be applied to this case. 
First, 
the Higgs boson obtains a thermal mass from hard $\chi_\phi$-particles just after inflation ($\ti \sim 1$) 
as $\alpha_\phi^{1/2} T_*$ with Eq.~(\ref{m_T^hard}), 
but this may be too small 
to restore the electroweak symmetry.
Then, 
at the time around $\ti = \ti_{\rm max}$, 
the effective temperature of the soft sector reaches a maximal value 
and the Higgs boson obtains a thermal mass from $\chi$-particles as $\alpha_\phi^{1/2} T_{*, \chi_\phi}$ with Eq.~(\ref{m_T for Higgs}). 
This can be
of order $100 \GEV$ for the case of $m_I = 10^{13} \GEV$
if the reheating temperature is slightly larger than $1 \MEV$, say $T_\text{RH} \simeq 5 \MEV$,
because $\alpha_\phi \approx y_t^2 / 4 \pi \sim 0.1$ and $\alpha \approx \alpha_s \sim 0.1$, 
where $y_t$ and $\alpha_s$ are the top Yukawa coupling and the strong coupling, respectively.\footnote{
	Although the top quark is fermion, its effective temperature 
	at a time around $\ti \sim \ti_{\rm max}$ coincides with the one for bosons, 
	which we calculate in this paper. 
}
Thus, 
the electroweak symmetry may be restored at the time around $\ti = \ti_{\rm max}$ 
even if the reheating temperature is as low as ${\cal O} (1) \MEV$ and the inflaton mass is $10^{13} \GEV$. 
When the Higgs field stays at the symmetric phase, 
the sphaleron effect, which is a $B+L$ violating process in a finite temperature plasma, 
is turned on~\cite{Kuzmin:1985mm}. 
Our result implies that there may be an era in which the sphaleron effect proceeds efficiently 
even if the reheating temperature is ${\cal O}(1) \MEV$. 
This is an important result 
when one determines the baryon asymmetry of the Universe
in such a low reheating temperature scenario.

\subsection{\label{AD field}Affleck-Dine field}

In SUSY theories, 
there are many scalar fields whose potentials are absent in the limit of exact SUSY and 
within the renormalizable level. 
Focusing on a baryonic scalar field with such a flat potential, called the AD field, 
baryon asymmetry can be generated by the Affleck-Dine mechanism~\cite{Affleck:1984fy, Dine:1995kz}. 
In this subsection, we identify $\phi$ as the AD field. 

The AD field $\phi$ has a large VEV due to a negative Hubble-induced mass during inflation 
and the inflaton oscillation dominated era. 
When the Hubble parameter becomes comparable to the curvature of the potential, 
it starts to oscillate around the origin of the potential. 
At the same time, it is kicked in the phase direction and generate baryon asymmetry.
The resulting baryon-to-entropy ratio $Y_b$ 
can be calculated from 
\begin{align}
	 Y_b \sim \frac{m_{3/2} T_{\rm RH}}{H_{\rm osc}^2} \lmk \frac{\abs{\phi_{\rm osc}}}{\Mpl} \rmk^2, 
	 \label{Y_b}
\end{align}
where we neglect $O(1)$ numerical factors (see, {\it e.g.}, Ref.~\cite{Harigaya:2014tla}). 
Here, $m_{3/2}$ is gravitino mass, $\phi_{\rm osc}$ is the VEV of $\phi$ at the time of beginning of its oscillation. 
Note that the amount of baryon asymmetry depends on the Hubble parameter at the beginning of oscillation $H_{\rm osc}$. 
Thus, we should include thermal effects on the potential of the AD field 
to determine $H_{\rm osc}$ as 
\begin{align}
	 H_{\rm osc}^2 \simeq \Max \lkk m_\phi^2, \ \alpha_s^2 \frac{T^4}{\abs{\phi_{\rm osc}}^2} \rkk, 
	 \label{H_osc}
\end{align}
where we assume $|g_\phi \phi | \gg T$~\cite{Fujii:2001zr, Anisimov:2000wx}. 
In the literature, they use the relation of $T = ( T_{\rm RH}^2 \Mpl H)^{1/4}$ 
to calculate $H_{\rm osc}$ [see Eq.~(\ref{T before reheating})]. 
However, as shown in this paper, the finite time scale of thermalization implies that 
this relation holds only after the time of $\ti = \ti_{\rm max}$, 
so that we should check that $H_{\rm osc}^{-1} m_\phi \gtrsim \ti_{\rm max}$ is satisfied.

Assuming $T = ( T_{\rm RH}^2 \Mpl H)^{1/4}$, we can rewrite Eq.~(\ref{H_osc}) as 
\begin{align}
	 H_{\rm osc} \simeq \Max \lkk m_\phi, \ \alpha_s^2 \frac{T_{\rm RH}^2 \Mpl}{\abs{\phi_{\rm osc}}^2} \rkk.
\end{align}
We require the following condition to use the relation of $T = ( T_{\rm RH}^2 \Mpl H)^{1/4}$: 
\begin{align} 
	 H_{\rm osc} \lesssim \alpha^{16/5} \Gphi^{3/5} m_I.
\end{align}
Using Eq.~(\ref{Y_b}) to eliminate $\abs{\phi}_{\rm osc}$ 
and assuming $H_{\rm osc} \simeq \alpha_s^2 T_{\rm RH}^2 \Mpl / \abs{\phi_{\rm osc}}^2$, 
we can rewrite the condition as 
\begin{align}
	 T_{\rm RH} \gtrsim 4 \GEV 
	 \times 
	 \lmk \frac{\alpha}{0.1} \rmk^{-38/3} 
	 \lmk \frac{Y_b}{10^{-10}} \rmk^{-5/3} 
	 \lmk \frac{m_I}{10^{13} \GEV} \rmk^{4} 
	 \lmk \frac{m_{3/2}}{1 \TEV} \rmk^{5/3}.
\end{align}
This is easily satisfied, 
so that we can consistently use the relation of $T = ( T_{\rm RH}^2 \Mpl H)^{1/4}$ 
to calculate the baryon asymmetry.

Finally, we comment on the case of 
``direct $\chi_\phi$-production''. 
When the VEV of the AD field is smaller than the inflaton mass, 
$\chi_\phi$-particles may be produced directly from the decay of inflaton. 
In this case, 
the AD field may obtain a large effective mass as Eq.~(\ref{m_T^hard}) just after inflation. 
Since this thermal mass easily exceeds the mass of the AD field $m_\phi$, 
calculations of the baryon asymmetry may be completely changed in this case. 
We leave this issue for a future work.

\section{\label{conclusion}Conclusion}

We have investigated the reheating process after inflation, 
and derived the evolutions of distribution function of light particles ($\chi$-particles),
and also effective temperatures that describe thermal effects on scalar fields. 
Our results are summarized in Sec.~\ref{summary in sec 3} 
and are illustrated as the blue lines in Figs.~\ref{fig1} and \ref{fig2}. 
Comparing with the result calculated by assuming the ``{\it instantaneous thermalization}'', 
which is widely assumed in the literature, 
we find that the maximal temperature of the Universe is overestimated in the literature,
in particular, for the case of $\alpha^3 \gtrsim \Gphi = \Gamma_I / (m_I^3 / \Mpl^2)$.
This is mainly because  
emissions of high momentum particles are necessary for primary particles to be thermalized
but such processes are suppressed by the LPM effect in reality.

We have applied the result to some important phenomena: 
the SSB of PQ symmetry, the electroweak symmetry breaking, 
and the dynamics of an AD field. 
We have found that 
the maximum temperature of the Universe can be much lower than
the conventional estimation where the ``{\it instantaneous thermalization}'' is assumed. 
In particular, we have shown that
if the reheating temperature is as low as ${\cal O} (1) \MEV$,
then the maximum temperature of the Universe may be at most electroweak scale.
This implies that 
the electroweak symmetry may be marginally restored after inflation 
even for the case of such a low reheating temperature. 
The PQ symmetry might not be restored after inflation in the axion cold dark matter scenario,
even if the reheating temperature is as large as the one required by the realization of thermal leptogenesis. 
Our results are applied to the calculation of the baryon asymmetry 
in the Affleck-Dine baryogenesis. 
We have justified the conventional calculations performed in the literature 
when the VEV of the AD field is larger than the mass of inflaton. 
Also, our result  implies that DM may not be able to be produced from the thermal plasma 
in a low reheating temperature scenario,
contrary to the conventional studies under the ``{\it instantaneous thermalization}'' assumption.

We have studied in detail the evolution of distribution function of radiation,
and how particles with masses of $|g_\phi \phi|$ are produced from radiation.
Thus, our results are also applicable to heavy particle production
in the detailed process of thermalization,
as partly done in Ref.~\cite{Harigaya:2014waa}.

\section*{Acknowledgments}

K.M. would like to thank M.~Takeuchi and S.~Matsumoto for helpful discussions.
This work was supported by World Premier International Research Center Initiative (WPI Initiative), MEXT, Japan. 
The work of K.M. and M.Y. was supported in part by JSPS Research Fellowships for Young Scientists, No. 25.8715 (M.Y.).
M.Y. acknowledges the support by the Program for Leading Graduate Schools, MEXT, Japan.

\appendix
\section{LPM-suppressed Splitting Rate}
\label{app:lpm}

Here we summarize basic formulae of LPM-suppressed splitting rate
in SU$(N)$ gauge theory with $N_F$-flavor fundamental fermions
to understand the model dependence of splitting rate.
We follow Ref.~\cite{Arnold:2002zm} in the following.

Let us first start with a precise form of the collision term for the splitting process.
When the primaries have larger energy than the typical screening mass scale of plasma,
like in our case, we have to take into account nearly collinear emissions.
The collision term in this case can be expressed as
\begin{align}
	{\cal C}_\text{split} [f_a] = & \quad
	\frac{\prn{2 \pi}^3}{2 k^2 \nu_a} \sum_{b,c}
	\int \dd p' \dd k' \delta \prn{k - p' - k'} \gamma^a_{bc} \prn{k; p', k'} \nonumber \\
	&\quad \quad \quad \quad \quad  
	\times \prn{
		f_a (k) \com{ 1 \pm f_b (p') } \com{ 1 \pm f_c (k') } - 
		\com{1 \pm f_a (k) } f_b (p') f_c (k')
	} \nonumber \\[.5em]
	&+ \frac{\prn{2 \pi}^3}{k^2 \nu_a} \sum_{b,c}
	\int \dd p' \dd p \delta \prn{k + p - p'} \gamma^c_{ab} \prn{p'; k, p} \nonumber \\
	&\quad \quad \quad \quad \quad  
	\times \prn{
		f_a (k) f_b (p) \com{ 1 \pm f_c (p') } - 
		\com{1 \pm f_a (k) } \com{ 1 \pm f_b (p)} f_c (p')
	},
	\label{eq:split_col_app}
\end{align}
where $\nu_a$ represents the number of degree of freedom for species $a$
(normalized by one real d.o.f.).
The splitting/joining process for $a \leftrightarrow bc$ is characterized by a splitting function
$\gamma^a_{bc}$.
As explained intuitively in Sec.~\ref{thermalization},
we have to include destructive interference effect, LPM effect,
in order to correctly cope with the collinear splitting process in medium.
The required splitting function which deals with the LPM effect at the leading logarithmic order 
is given by~\cite{Arnold:2008zu,Arnold:2002zm}
\begin{align}
	\gamma^q_{q g} \prn{P; \prn{1 - x}P, x P}
	&= \gamma^{\bar q}_{\bar q g} \prn{P; \prn{1 - x}P, x P} \nonumber \\
	&= \frac{d_F C_F \alpha}{4 \pi^3 \sqrt{2}} 
	\com{ \hat q_\text{el} P_\text{min} }^\frac{1}{2} \frac{1 + \prn{1 - x}^2}{x^2 \prn{1- x}}
	\times LL \prn{x; F, A, F}, \\[.5em]
	\gamma^g_{q \bar q} \prn{P; x P, \prn{1 - x}P}
	&= \frac{d_F C_F \alpha}{ 4 \pi^3 \sqrt{2}}
	\com{ \hat q_\text{el} P_\text{min} }^\frac{1}{2} \frac{x^2 + \prn{1 - x}^2}{x \prn{1- x}}
	\times LL\prn{x; A, F, F }, \\[.5em]
	\gamma^g_{g g} \prn{P; x P, \prn{1 - x}P}
	&= \frac{d_A C_A \alpha}{ 4 \pi^3 \sqrt{2}}
	\com{ \hat q_\text{el} P_\text{min} }^\frac{1}{2} \frac{1 + x^4 + \prn{1 - x}^4}{x^2 \prn{1- x}^2}
	\times LL \prn{x; A, A, A },
\end{align}
where $P_\text{min} \equiv \Min \com{ 1, \prn{1 - x} } P$, 
$d_a$ is the dimension of representation for a species $a$,
and $C_a$ is the quadratic Casimirs for a species $a$.
$q$ and $g$ represent a fermion with the fundamental representation
and a gauge boson of SU$(N)$ respectively.
The diffusion coefficient is defined as
\begin{align}
	\hat q_\text{el}
	\equiv \alpha^2 \sum_a \bar \nu_a t_a \int_{\bm{p}} f_a (\bm{p}) \com{ 1 \pm f_a (\bm{p}) },
	\label{eq:diff_coeff}
\end{align}
with $\bar \nu_a \equiv \nu_a / d_a$ being the number of d.o.f. of a species $a$ excluding
the ``color'' factor and $t_a$ being the normalization of representation for a species $a$.
The Leading Logarithmic enhancement factor is given by
\begin{align}
	LL \prn{x; s_1, s_2, s_3} 
	= & \quad \com{ \frac{2}{\pi} \Max \com{ x, 1- x } }^\frac{1}{2} \nonumber \\
	&\times \com{
		\frac{1}{2} \prn{ C_{s_2} + C_{s_3} - C_{s_1}}
		+ \frac{1}{2} \prn{ C_{s_3} + C_{s_1} - C_{s_2}} x^2
		+ \frac{1}{2} \prn{ C_{s_1} + C_{s_2} - C_{s_3}} \prn{1 - x}^2
	}^\frac{1}{2} \nonumber \\
	& \times \com{ \ln \prn{ \frac{P}{T_\text{eff}} } }^\frac{1}{2},
\end{align}
where the effective temperature is defined as
\begin{align}
	T_\text{eff} \equiv \frac{\sum_a \bar \nu_a t_a \int_{\bm{p}} f_a (\bm{p}) 
	\com{ 1 \pm f_a (\bm{p})} }
	{2 \sum_a  \bar \nu_a t_a \int_{\bm{p}}\frac{f_a (\bm{p})}{p}}.
\end{align}
For instance, for a thermal-like distribution $\sim T_s / p$, like Eq.~\eqref{f_s thermal 2}, 
the effective temperature, $T_\text{eff}$, coincides with the soft temperature, $T_s$.

It is instructive to reproduce the important equation [Eq.~\eqref{f_s}],
which is responsible for  production of the soft population from hard primaries,
and also determines $\tilde t_\text{max}$ and hence $T_\text{max}$.
To be concrete, let us assume that inflaton decays directly into SM gauge bosons.
The Boltzmann equation responsible for production of soft particles is given by
\begin{align}
	\frac{1}{\prn{2 \pi}^3} \dot f_{g,s}(t, k) 
	&= \int \dd p' \dd p \delta \prn{ k + p  - p'} 
	\frac{1}{k^2 \nu_g} \gamma^g_{gg} (p'; k, p)
	f_{g,h} (p') + \cdots\\
	&\simeq
	\frac{\sqrt{2} C_A}{\bar \nu_g}
	LL \prn{ 0; A,A,A }  \times  \alpha \com{ \frac{\hat q_\text{el}}{k} }^\frac{1}{2}
	\times n_h k^{-3}.
\end{align}
Thus, the splitting rate can be parametrized as
\begin{align}
	\Gamma_\text{split} (k)
	= c' \alpha \com{ \frac{\hat q_\text{el}}{k} }^\frac{1}{2}~~~\text{with}~~
	c' \equiv \frac{\sqrt{2} C_A}{\bar \nu_g} LL (0; A, A, A).
\end{align}
The coefficient $c'$ is given by a model dependent factor times the leading log enhancement.
One can derive a similar formula when the inflaton decays into fermions 
and then it emits gauge bosons collinearly.
Also, Eq.~\eqref{eq:split_apprx} can be derived from Eq.~\eqref{eq:split_col_app}.

In the main part of this paper, we have omitted this factor to avoid model dependent discussions,
but one can see that this factor results in at most ${\cal O} (10)$ uncertainty.
As discussed in the end of Sec.~\ref{thermalization},
the maximum temperature has a mild dependence on this uncertainty factor, $\propto c'^{2/5}$.
Together with the numerical factor derived from numerical calculations of Boltzmann equation,
we conservatively conclude that the uncertainty of maximum temperature is at most a factor of ten.

\section{Effective Temperature for $\chi_\phi$-particles}
\label{app:eff_t}

In contrast to $\chi$-particles, $\chi_\phi$-particles can have sizable masses of $|g_\phi \phi|$
due to the large expectation value of the $\phi$-condensate.
The distribution in momentum space can be different from that of $\chi$-particles and hence separate discussions are required.
In this Appendix, we 
show the results of $T_{*, \chi_\phi}$.

\subsection{$\ti_{\rm ini} < \ti < \ti_{\rm soft}$}

If the effective mass of $\chi_\phi$-particles are smaller than the screening mass, 
then the effective temperature for $\chi_\phi$-particles
coincide with that for $\chi$-particles: 
\begin{align}
	\left. T_{\ast, \chi_\phi}^2  \right|_\text{indir} \sim T_\ast^2,
	\label{eq:effT_small_1}
\end{align}
for $| g_\phi \phi | < m_s$.
Here the subscript ``indir'' indicates $\chi_\phi$-particles which are not produced directly from inflaton decay.
For the case of $m_s < |g_\phi \phi|$, 
the effective temperature squared $T_{\ast, \chi_\phi}^2$ 
is evaluated as\footnote{
	Here we neglect $t$-channel contributions in the soft sector, 
	because they are always subdominant to calculate the thermal potential. 
}
\begin{align}
	\left. T_{\ast, \chi_\phi}^2 \right|_\text{indir} \equiv \Max \left[ \left. T_{\ast, \chi_\phi}^2 \right|_\text{hard},
	\left. T_{\ast, \chi_\phi}^2 \right|_\text{soft},
	\left. T_{\ast, \chi_\phi}^2 \right|_\text{int} \right],
	\label{eq:effT_large_1}
\end{align}
where
\begin{align}
	\left. T_{\ast, \chi_\phi}^2 \right|_\text{hard}
	& \sim  
	T_\ast^2 \left( \frac{m_s}{|g_\phi \phi|} \right) \, \Min \left[ 1, \left( \frac{k_\text{max} m_s}{(g_\phi \phi)^2} \right) \right],
	\label{eq:effT_large_1_hard} \\[.5em]
	\left. T_{\ast, \chi_\phi}^2 \right|_\text{soft}
	& \sim  
	\alpha^2 \Gphi m_I^2 \, \Min \left[ 
	1, \left( \frac{k_\text{max}}{g_\phi \phi} \right)^4 \Max \com{1, \prn{ \frac{|g_\phi \phi|}{(k_\text{max} k_\text{form})^{1/2}} }} \right], \\[.5em]
	\left. T_{\ast, \chi_\phi}^2 \right|_\text{int}
	& \sim
	\alpha^2 \Gphi^{3/2} \ti^{-1/2} m_I^2 \,
	\Min \left[ 1, \left( \frac{k_\text{max} m_I}{ |g_\phi \phi|^2 } \right)^2 \Max \com{ 1, \prn{ \frac{|g_\phi \phi|}{(k_\text{form} m_I)^{1/2}} } } \right]. 
\end{align}
By using these equations,
one can show that, for $\alpha^{-1} k_\text{max} \lesssim |g_\phi \phi|$ with $\Gphi \lesssim \alpha^6$,
the contribution of abundant $\chi_\phi$-particles
is always subdominant compared with that of the running coupling constant.
Even in the case of $\Gphi \gtrsim \alpha^6$,
its effect is roughly the same order with that of the running coupling constant
for $|g_\phi \phi| \gtrsim \alpha^{-1} k_\text{max}$.
Hence,
one can omit the decay of massive $\chi_\phi$-particles in order to show that 
the effect on the effective potential of $\phi$ from $\chi_\phi$-particles is subdominant
for the large field value regime.

However, if $\chi_\phi$-particles are produced directly from inflaton decay, 
it has a sizable contribution to the effective temperature,
which tend to dominate for a large field value [see Eq.~\eqref{eq:p1_dir}].
Therefore, 
we should take into account the decay of $\chi_\phi$-particles 
to obtain more realistic predictions at least in that case:
\begin{align}
	\left. T_{\ast, \chi_\phi}^2 \right|_\text{dir}
	= \alpha \Gphi^{1/2} T_\ast^2 \left( \frac{\ti_\text{soft}}{\ti} \right)^{1/2}
	\Min \left[ 1, \frac{1}{\Gamma_\text{decay} t} \right],
	\label{eq:effT_large_1_i}
\end{align}
where the subscript ``dir'' indicates contribution from $\chi_\phi$-particles which are directly produced from inflaton decay.
Hence, one should compare it with the log potential from the running coupling constant.
The effective mass squared for the large field value regime, $|g_\phi \phi| \gtrsim \alpha^{-1} k_\text{max}$, should be given by
\begin{align}
	\Max \left[ \alpha^2 \frac{T_\ast^4}{\phi^2},~
	\left. \alpha_\phi T_{\ast, \chi_\phi}^2 \right|_\text{dir} \right].
\end{align}

\subsection{$\ti_{\rm soft} < \ti < \ti_{\rm max}$}

If the effective mass of $\chi_\phi$-particles are smaller than the screening mass, 
then the effective temperature for $\chi_\phi$-particles
coincide with that for $\chi$-particles: 
\begin{align}
	\left. T_{\ast, \chi_\phi}^2  \right|_\text{indir} \sim T_s^2,
	\label{eq:effT_small_1}
\end{align}
for $| g_\phi \phi | < m_s$.
Here the subscript ``indir'' indicates $\chi_\phi$-particles which are not produced directly from inflaton decay.
For the case of $m_s < |g_\phi \phi|$, 
we obtain the effective temperature squared $T_{\ast, \chi_\phi}^2$ 
as 
\begin{align}
	\left. T_{\ast, \chi_\phi}^2 \right|_\text{indir} \equiv \Max \left[ \left. T_{\ast, \chi_\phi}^2 \right|_\text{hard},
	\left. T_{\ast, \chi_\phi}^2 \right|_\text{soft},
	\left. T_{\ast, \chi_\phi}^2 \right|_\text{int} \right],
	\label{eq:effT_large_1}
\end{align}
where
\begin{align}
	\left. T_{\ast, \chi_\phi}^2 \right|_\text{hard}
	& \sim  
	T_s^2 \left( \frac{\ti_{\rm soft}}{\ti} \right)^3 \left( \frac{m_s}{|g_\phi \phi|} \right) \, \Min \left[ 1, \frac{T_s m_s}{(g_\phi \phi)^2} \left( \frac{\ti_{\rm soft}}{\ti} \right)^3 \right],
	\label{eq:effT_large_1_hard} \\[.5em]
	\left. T_{\ast, \chi_\phi}^2 \right|_\text{soft}
	& \sim  
	T_s^2
	\, \Min \left[ 
	1, \left( \frac{T_s}{g_\phi \phi} \right)^4 
		\Max \com{1, \prn{  \frac{|g_\phi \phi|}{( T_s k_\text{form} )^{1/2}} }} 
	\right], \\[.5em]
	\left. T_{\ast, \chi_\phi}^2 \right|_\text{int}
	& \sim
		\begin{cases}
		\Max \left[
		\alpha^{3/2} \Gphi^{3/4} T_s^2 \left( \cfrac{\ti_{\rm soft}}{\ti} \right)^{1/2}, 
		\alpha T_s^2 \left( \cfrac{\ti_{\rm soft}}{\ti} \right)^{2} \left( \cfrac{T_s}{|g_\phi \phi|} \right)^{3}
		\right] 
		\\[2em]& \!\!\!\!\!\!\!\!\!\!\!\!\!\!\!\! \!\!\!\!\!\!\!\!\!\!\!\!\!\!\!\! \!\!\!\!\!\!\!\!\!\!\!\!\!\!\!\! \!\!\!\!\!\!\!\!\!\!\!\!\!\!\!\! 
		\text{for}~~ (k_{\rm th} T_s)^{1/2} \lesssim |g_\phi \phi| \lesssim (T_s m_I)^{1/2}, \\[1em]
		\alpha^{3/2} \Gphi^{3/4} \left( \cfrac{\ti_{\rm soft}}{\ti} \right)^{5/2} T_s^2 
		\left( \cfrac{T_s m_I}{|g_\phi \phi|^2} \right)^2 \Max \com{ 1, \prn{ \cfrac{|g_\phi \phi|}{(m_I k_\text{form})^{1/2}} } }
		\\[2em]& \!\!\!\!\!\!\!\!\!\!\!\!\!\!\!\! \!\!\!\!\!\!\!\!\!\!\!\!\!\!\!\! \!\!\!\!\!\!\!\!\!\!\!\!\!\!\!\! \!\!\!\!\!\!\!\!\!\!\!\!\!\!\!\! 
		\text{for}~~ (T_s m_I)^{1/2} \lesssim |g_\phi \phi|, 
		\end{cases}
\end{align}
where $k_\text{th}$ represents the interval boundary the hard and soft sectors.
Here we include the contribution coming from $t$-channel scatterings between the hard and soft sector. 
Again, one can show that the effective potential from $\chi_\phi$-particles are always subdominant
for the case of $|g_\phi \phi| \gtrsim \alpha^{-1} T_\ast$ with $\Gphi \lesssim \alpha^{-6}$.
And also even in the case of $\Gphi \gtrsim \alpha^{-6}$, it is roughly the same order of that from the running coupling constant.
Thus, we can omit the decay of $\chi_\phi$-particles in order to show that it is always subdominant compared with that from
the running coupling constant for the large field regime.

However, if the inflaton directly decay into $\chi_\phi$-particles, then the effective temperature
from direct inflaton decay tends to dominate for the large field value, $|g_\phi \phi| \gtrsim \alpha^{-1} T_\ast$.\footnote{
	For the small field value regime of $|g_\phi \phi| \lesssim T_\ast$,
	one can show that the soft $\chi_\phi$-particles always dominate over that of hard ones produced via direct inflaton decay.
}
Hence, one should compare it with that from the running coupling constant:
\begin{align}
	\Max \left[ \alpha^2 \frac{T_{\ast}^4}{\phi^2},~
	\left. \alpha_\phi T_{\ast, \chi_\phi}^2 \right|_\text{dir}  \right],
\end{align}
where $T_{\ast, \chi_\phi}^2 |_\text{dir}$ is given by Eq.~\eqref{eq:effT_large_1_i} [See also Eq.~\eqref{eq:p2_dir}].

\subsection{$\ti_{\rm max} < \ti < \ti_{\rm RH}$}

Calculating the distribution of $\chi_\phi$-particles, 
we obtain the effective temperature squared $T_{\ast, \chi_\phi}^2$ 
as 
\begin{align}
	\left. T_{\ast, \chi_\phi}^2 \right|_\text{indir}
	& \sim
		\begin{cases}		
		T_\ast^2
		&\text{for}~~ |g_\phi \phi| \lesssim T_\ast \\[1.5em]		
		T_\ast^2 \lmk \cfrac{T_\ast}{|g_\phi \phi|} \rmk^4 
		&\text{for}~~ T_\ast \lesssim |g_\phi \phi| \lesssim (T_\ast m_I)^{1/2} \\[1.5em]		
		T_\ast^2 \alpha^2 \Gphi \left( \cfrac{\ti_{\rm max}}{\ti} \right)^{5/4}  
		\left( \cfrac{(T_\ast m_I)^{1/2}}{|g_\phi \phi|} \right)^4 
		&\text{for}~~ (T_\ast m_I)^{1/2} \lesssim |g_\phi \phi|, 
		\end{cases}
\end{align}
where the subscript ``indir'' indicates $\chi_\phi$-particles which are not produced directly from inflaton decay.
At that time, one can show that the contribution from abundant $\chi_\phi$-particles is always subdominant
compared with that from the running coupling constant for the large field value regime $|g_\phi \phi| \gtrsim \alpha^{-1} T_\ast$.
Thus, we can omit the decay of $\chi_\phi$-particles in order to show that it is always subdominant compared with that from
the running coupling constant for the large field regime.

However, if the inflaton directly decay into $\chi_\phi$-particles, 
then its contribution tends to dominate for the large field value regime, $|g_\phi \phi| \gtrsim \alpha^{-1} T_\ast$.
Hence, one should compare it with that from the running coupling constant:
\begin{align}
	\Max \left[ \alpha \frac{ T_{\ast}^4 }{\phi^2},~ \left. T_{\ast, \chi_\phi}^2 \right|_\text{dir}  \right],
\end{align}
where $T_{\ast, \chi_\phi}^2 |_\text{dir}$ is given by [See also Eq.~\eqref{eq:p3_dir}]
\begin{align}
	\left. T_{\ast, \chi_\phi}^2 \right|_\text{dir} \sim
	\Gphi \ti^{-1} m_I^2\, \Min \left[ \frac{1}{\Gamma_\text{split} (m_I) t}, \frac{1}{\Gamma_\text{decay} t} \right].
\end{align}
Here note that the hard particles with the momentum of $m_I$ soon breaks up within the time scale of $\Gamma_\text{split}$,
and hence $1/ [\Gamma_\text{split} (m_I) t]$ should be multiplied.

\bibliography{ref}

\providecommand{\href}[2]{#2}\begingroup\raggedright\begin{thebibliography}{10}

\bibitem{Linde:1981mu}
A.~D. Linde, ``{A New Inflationary Universe Scenario: A Possible Solution of
  the Horizon, Flatness, Homogeneity, Isotropy and Primordial Monopole
  Problems},''
\href{http://dx.doi.org/10.1016/0370-2693(82)91219-9}{{\em Phys.Lett.}
  {\bfseries B108} (1982) 389--393}.

\bibitem{Albrecht:1982mp}
A.~Albrecht, P.~J. Steinhardt, M.~S. Turner, and F.~Wilczek, ``{Reheating an
  Inflationary Universe},''
\href{http://dx.doi.org/10.1103/PhysRevLett.48.1437}{{\em Phys.Rev.Lett.}
  {\bfseries 48} (1982) 1437}.

\bibitem{Weinberg:1977ma}
S.~Weinberg, ``{A New Light Boson?},''
\href{http://dx.doi.org/10.1103/PhysRevLett.40.223}{{\em Phys.Rev.Lett.}
  {\bfseries 40} (1978) 223--226}.

\bibitem{Peccei:1977hh}
R.~Peccei and H.~R. Quinn, ``{CP Conservation in the Presence of Instantons},''
\href{http://dx.doi.org/10.1103/PhysRevLett.38.1440}{{\em Phys.Rev.Lett.}
  {\bfseries 38} (1977) 1440--1443}.

\bibitem{Peccei:1977ur}
R.~Peccei and H.~R. Quinn, ``{Constraints Imposed by CP Conservation in the
  Presence of Instantons},''
\href{http://dx.doi.org/10.1103/PhysRevD.16.1791}{{\em Phys.Rev.} {\bfseries
  D16} (1977) 1791--1797}.

\bibitem{Affleck:1984fy}
I.~Affleck and M.~Dine, ``{A New Mechanism for Baryogenesis},''
\href{http://dx.doi.org/10.1016/0550-3213(85)90021-5}{{\em Nucl.Phys.}
  {\bfseries B249} (1985) 361}.

\bibitem{Dine:1995kz}
M.~Dine, L.~Randall, and S.~D. Thomas, ``{Baryogenesis from flat directions of
  the supersymmetric standard model},''
  \href{http://dx.doi.org/10.1016/0550-3213(95)00538-2}{{\em Nucl.Phys.}
  {\bfseries B458} (1996) 291--326},
\href{http://arxiv.org/abs/hep-ph/9507453}{{\ttfamily arXiv:hep-ph/9507453
  [hep-ph]}}.

\bibitem{Yokoyama:2004pf}
J.~Yokoyama, ``{Fate of oscillating scalar fields in the thermal bath and their
  cosmological implications},''
  \href{http://dx.doi.org/10.1103/PhysRevD.70.103511}{{\em Phys.Rev.}
  {\bfseries D70} (2004) 103511},
\href{http://arxiv.org/abs/hep-ph/0406072}{{\ttfamily arXiv:hep-ph/0406072
  [hep-ph]}}.

\bibitem{Mukaida:2012qn}
K.~Mukaida and K.~Nakayama, ``{Dynamics of oscillating scalar field in thermal
  environment},'' \href{http://dx.doi.org/10.1088/1475-7516/2013/01/017}{{\em
  JCAP} {\bfseries 1301} (2013) 017},
\href{http://arxiv.org/abs/1208.3399}{{\ttfamily arXiv:1208.3399 [hep-ph]}}.

\bibitem{Mukaida:2012bz}
K.~Mukaida and K.~Nakayama, ``{Dissipative Effects on Reheating after
  Inflation},'' \href{http://dx.doi.org/10.1088/1475-7516/2013/03/002}{{\em
  JCAP} {\bfseries 1303} (2013) 002},
\href{http://arxiv.org/abs/1212.4985}{{\ttfamily arXiv:1212.4985 [hep-ph]}}.

\bibitem{Mukaida:2013xxa}
K.~Mukaida, K.~Nakayama, and M.~Takimoto, ``{Fate of $Z_2$ Symmetric Scalar
  Field},'' \href{http://dx.doi.org/10.1007/JHEP12(2013)053}{{\em JHEP}
  {\bfseries 1312} (2013) 053},
\href{http://arxiv.org/abs/1308.4394}{{\ttfamily arXiv:1308.4394 [hep-ph]}}.

\bibitem{Drewes:2013iaa}
M.~Drewes and J.~U. Kang, ``{The Kinematics of Cosmic Reheating},''
  \href{http://dx.doi.org/10.1016/j.nuclphysb.2013.07.009,
  10.1016/j.nuclphysb.2014.09.008}{{\em Nucl.Phys.} {\bfseries B875} (2013)
  315--350},
\href{http://arxiv.org/abs/1305.0267}{{\ttfamily arXiv:1305.0267 [hep-ph]}}.

\bibitem{Cheung:2015iqa}
Y.-K.~E. Cheung, M.~Drewes, J.~U. Kang, and J.~C. Kim, ``{Effective Action for
  Cosmological Scalar Fields at Finite Temperature},''
\href{http://arxiv.org/abs/1504.04444}{{\ttfamily arXiv:1504.04444 [hep-ph]}}.

\bibitem{Kirzhnits:1972iw}
D.~Kirzhnits, ``{Weinberg model in the hot universe},''
{\em JETP Lett.} {\bfseries 15} (1972) 529--531.

\bibitem{Kirzhnits:1972ut}
D.~Kirzhnits and A.~D. Linde, ``{Macroscopic Consequences of the Weinberg
  Model},''
\href{http://dx.doi.org/10.1016/0370-2693(72)90109-8}{{\em Phys.Lett.}
  {\bfseries B42} (1972) 471--474}.

\bibitem{Dolan:1973qd}
L.~Dolan and R.~Jackiw, ``{Symmetry Behavior at Finite Temperature},''
\href{http://dx.doi.org/10.1103/PhysRevD.9.3320}{{\em Phys.Rev.} {\bfseries D9}
  (1974) 3320--3341}.

\bibitem{Weinberg:1974hy}
S.~Weinberg, ``{Gauge and Global Symmetries at High Temperature},''
\href{http://dx.doi.org/10.1103/PhysRevD.9.3357}{{\em Phys.Rev.} {\bfseries D9}
  (1974) 3357--3378}.

\bibitem{Kirzhnits:1976ts}
D.~Kirzhnits and A.~D. Linde, ``{Symmetry Behavior in Gauge Theories},''
\href{http://dx.doi.org/10.1016/0003-4916(76)90279-7}{{\em Annals Phys.}
  {\bfseries 101} (1976) 195--238}.

\bibitem{Zeldovich:1974uw}
Y.~Zeldovich, I.~Y. Kobzarev, and L.~Okun, ``{Cosmological Consequences of the
  Spontaneous Breakdown of Discrete Symmetry},''
{\em Zh.Eksp.Teor.Fiz.} {\bfseries 67} (1974) 3--11.

\bibitem{Sikivie:1982qv}
P.~Sikivie, ``{Of Axions, Domain Walls and the Early Universe},''
\href{http://dx.doi.org/10.1103/PhysRevLett.48.1156}{{\em Phys.Rev.Lett.}
  {\bfseries 48} (1982) 1156--1159}.

\bibitem{Axenides:1983hj}
M.~Axenides, R.~H. Brandenberger, and M.~S. Turner, ``{Development of Axion
  Perturbations in an Axion Dominated Universe},''
\href{http://dx.doi.org/10.1016/0370-2693(83)90586-5}{{\em Phys.Lett.}
  {\bfseries B126} (1983) 178}.

\bibitem{Seckel:1985tj}
D.~Seckel and M.~S. Turner, ``{Isothermal Density Perturbations in an Axion
  Dominated Inflationary Universe},''
\href{http://dx.doi.org/10.1103/PhysRevD.32.3178}{{\em Phys.Rev.} {\bfseries
  D32} (1985) 3178}.

\bibitem{Turner:1990uz}
M.~S. Turner and F.~Wilczek, ``{Inflationary axion cosmology},''
\href{http://dx.doi.org/10.1103/PhysRevLett.66.5}{{\em Phys.Rev.Lett.}
  {\bfseries 66} (1991) 5--8}.

\bibitem{Allahverdi:2000zd}
R.~Allahverdi, B.~A. Campbell, and J.~R. Ellis, ``{Reheating and supersymmetric
  flat direction baryogenesis},''
  \href{http://dx.doi.org/10.1016/S0550-3213(00)00124-3}{{\em Nucl.Phys.}
  {\bfseries B579} (2000) 355--375},
\href{http://arxiv.org/abs/hep-ph/0001122}{{\ttfamily arXiv:hep-ph/0001122
  [hep-ph]}}.

\bibitem{Asaka:2000nb}
T.~Asaka, M.~Fujii, K.~Hamaguchi, and T.~Yanagida, ``{Affleck-Dine leptogenesis
  with an ultralight neutrino},''
  \href{http://dx.doi.org/10.1103/PhysRevD.62.123514}{{\em Phys.Rev.}
  {\bfseries D62} (2000) 123514},
\href{http://arxiv.org/abs/hep-ph/0008041}{{\ttfamily arXiv:hep-ph/0008041
  [hep-ph]}}.

\bibitem{Fujii:2001zr}
M.~Fujii, K.~Hamaguchi, and T.~Yanagida, ``{Reheating temperature independence
  of cosmological baryon asymmetry in Affleck-Dine leptogenesis},''
  \href{http://dx.doi.org/10.1103/PhysRevD.63.123513}{{\em Phys.Rev.}
  {\bfseries D63} (2001) 123513},
\href{http://arxiv.org/abs/hep-ph/0102187}{{\ttfamily arXiv:hep-ph/0102187
  [hep-ph]}}.

\bibitem{Anisimov:2000wx}
A.~Anisimov and M.~Dine, ``{Some issues in flat direction baryogenesis},''
  \href{http://dx.doi.org/10.1016/S0550-3213(01)00550-8}{{\em Nucl.Phys.}
  {\bfseries B619} (2001) 729--740},
\href{http://arxiv.org/abs/hep-ph/0008058}{{\ttfamily arXiv:hep-ph/0008058
  [hep-ph]}}.

\bibitem{Chung:1998rq}
D.~J. Chung, E.~W. Kolb, and A.~Riotto, ``{Production of massive particles
  during reheating},'' \href{http://dx.doi.org/10.1103/PhysRevD.60.063504}{{\em
  Phys.Rev.} {\bfseries D60} (1999) 063504},
\href{http://arxiv.org/abs/hep-ph/9809453}{{\ttfamily arXiv:hep-ph/9809453
  [hep-ph]}}.

\bibitem{Giudice:2000ex}
G.~F. Giudice, E.~W. Kolb, and A.~Riotto, ``{Largest temperature of the
  radiation era and its cosmological implications},''
  \href{http://dx.doi.org/10.1103/PhysRevD.64.023508}{{\em Phys.Rev.}
  {\bfseries D64} (2001) 023508},
\href{http://arxiv.org/abs/hep-ph/0005123}{{\ttfamily arXiv:hep-ph/0005123
  [hep-ph]}}.

\bibitem{Harigaya:2013vwa}
K.~Harigaya and K.~Mukaida, ``{Thermalization after/during Reheating},''
  \href{http://dx.doi.org/10.1007/JHEP05(2014)006}{{\em JHEP} {\bfseries 1405}
  (2014) 006},
\href{http://arxiv.org/abs/1312.3097}{{\ttfamily arXiv:1312.3097 [hep-ph]}}.

\bibitem{Davidson:2000er}
S.~Davidson and S.~Sarkar, ``{Thermalization after inflation},''
  \href{http://dx.doi.org/10.1088/1126-6708/2000/11/012}{{\em JHEP} {\bfseries
  0011} (2000) 012},
\href{http://arxiv.org/abs/hep-ph/0009078}{{\ttfamily arXiv:hep-ph/0009078
  [hep-ph]}}.

\bibitem{Allahverdi:2002pu}
R.~Allahverdi and M.~Drees, ``{Thermalization after inflation and production of
  massive stable particles},''
  \href{http://dx.doi.org/10.1103/PhysRevD.66.063513}{{\em Phys.Rev.}
  {\bfseries D66} (2002) 063513},
\href{http://arxiv.org/abs/hep-ph/0205246}{{\ttfamily arXiv:hep-ph/0205246
  [hep-ph]}}.

\bibitem{Harigaya:2014waa}
K.~Harigaya, M.~Kawasaki, K.~Mukaida, and M.~Yamada, ``{Dark Matter Production
  in Late Time Reheating},''
  \href{http://dx.doi.org/10.1103/PhysRevD.89.083532}{{\em Phys.Rev.}
  {\bfseries D89} no.~8, (2014) 083532},
\href{http://arxiv.org/abs/1402.2846}{{\ttfamily arXiv:1402.2846 [hep-ph]}}.

\bibitem{Kofman:1994rk}
L.~Kofman, A.~D. Linde, and A.~A. Starobinsky, ``{Reheating after inflation},''
  \href{http://dx.doi.org/10.1103/PhysRevLett.73.3195}{{\em Phys.Rev.Lett.}
  {\bfseries 73} (1994) 3195--3198},
\href{http://arxiv.org/abs/hep-th/9405187}{{\ttfamily arXiv:hep-th/9405187
  [hep-th]}}.

\bibitem{Kofman:1997yn}
L.~Kofman, A.~D. Linde, and A.~A. Starobinsky, ``{Towards the theory of
  reheating after inflation},''
  \href{http://dx.doi.org/10.1103/PhysRevD.56.3258}{{\em Phys.Rev.} {\bfseries
  D56} (1997) 3258--3295},
\href{http://arxiv.org/abs/hep-ph/9704452}{{\ttfamily arXiv:hep-ph/9704452
  [hep-ph]}}.

\bibitem{McDonald:1999hd}
J.~McDonald, ``{Reheating temperature and inflaton mass bounds from
  thermalization after inflation},''
  \href{http://dx.doi.org/10.1103/PhysRevD.61.083513}{{\em Phys.Rev.}
  {\bfseries D61} (2000) 083513},
\href{http://arxiv.org/abs/hep-ph/9909467}{{\ttfamily arXiv:hep-ph/9909467
  [hep-ph]}}.

\bibitem{Moroi:2013tea}
T.~Moroi, K.~Mukaida, K.~Nakayama, and M.~Takimoto, ``{Scalar Trapping and
  Saxion Cosmology},'' \href{http://dx.doi.org/10.1007/JHEP06(2013)040}{{\em
  JHEP} {\bfseries 1306} (2013) 040},
\href{http://arxiv.org/abs/1304.6597}{{\ttfamily arXiv:1304.6597 [hep-ph]}}.

\bibitem{Kurkela:2011ti}
A.~Kurkela and G.~D. Moore, ``{Thermalization in Weakly Coupled Nonabelian
  Plasmas},'' \href{http://dx.doi.org/10.1007/JHEP12(2011)044}{{\em JHEP}
  {\bfseries 1112} (2011) 044},
\href{http://arxiv.org/abs/1107.5050}{{\ttfamily arXiv:1107.5050 [hep-ph]}}.

\bibitem{Kurkela:2014tea}
A.~Kurkela and E.~Lu, ``{Approach to Equilibrium in Weakly Coupled Non-Abelian
  Plasmas},'' \href{http://dx.doi.org/10.1103/PhysRevLett.113.182301}{{\em
  Phys.Rev.Lett.} {\bfseries 113} no.~18, (2014) 182301},
\href{http://arxiv.org/abs/1405.6318}{{\ttfamily arXiv:1405.6318 [hep-ph]}}.

\bibitem{baym1962quantum}
G.~Baym and L.~P. Kadanoff, {\em Quantum statistical mechanics}, vol.~1.
\newblock WA Benjamin, New York, 1962.

\bibitem{Calzetta:1986cq}
E.~Calzetta and B.~L. Hu, ``{Nonequilibrium Quantum Fields: Closed Time Path
  Effective Action, Wigner Function and Boltzmann Equation},''
\href{http://dx.doi.org/10.1103/PhysRevD.37.2878}{{\em Phys. Rev.} {\bfseries
  D37} (1988) 2878}.

\bibitem{Blaizot:2001nr}
J.-P. Blaizot and E.~Iancu, ``{The Quark gluon plasma: Collective dynamics and
  hard thermal loops},''
  \href{http://dx.doi.org/10.1016/S0370-1573(01)00061-8}{{\em Phys. Rept.}
  {\bfseries 359} (2002) 355--528},
\href{http://arxiv.org/abs/hep-ph/0101103}{{\ttfamily arXiv:hep-ph/0101103
  [hep-ph]}}.

\bibitem{Berges:2004yj}
J.~Berges, ``{Introduction to nonequilibrium quantum field theory},''
  \href{http://dx.doi.org/10.1063/1.1843591}{{\em AIP Conf. Proc.} {\bfseries
  739} (2005) 3--62}, \href{http://arxiv.org/abs/hep-ph/0409233}{{\ttfamily
  arXiv:hep-ph/0409233 [hep-ph]}}.
[,3(2004)].

\bibitem{Landau:1953um}
L.~Landau and I.~Pomeranchuk, ``{Limits of applicability of the theory of
  bremsstrahlung electrons and pair production at high-energies},''
{\em Dokl.Akad.Nauk Ser.Fiz.} {\bfseries 92} (1953) 535--536.

\bibitem{Migdal:1956tc}
A.~B. Migdal, ``{Bremsstrahlung and pair production in condensed media at
  high-energies},''
\href{http://dx.doi.org/10.1103/PhysRev.103.1811}{{\em Phys.Rev.} {\bfseries
  103} (1956) 1811--1820}.

\bibitem{Gyulassy:1993hr}
M.~Gyulassy and X.-n. Wang, ``{Multiple collisions and induced gluon
  Bremsstrahlung in QCD},''
  \href{http://dx.doi.org/10.1016/0550-3213(94)90079-5}{{\em Nucl.Phys.}
  {\bfseries B420} (1994) 583--614},
\href{http://arxiv.org/abs/nucl-th/9306003}{{\ttfamily arXiv:nucl-th/9306003
  [nucl-th]}}.

\bibitem{Arnold:2001ba}
P.~B. Arnold, G.~D. Moore, and L.~G. Yaffe, ``{Photon emission from
  ultrarelativistic plasmas},''
  \href{http://dx.doi.org/10.1088/1126-6708/2001/11/057}{{\em JHEP} {\bfseries
  0111} (2001) 057},
\href{http://arxiv.org/abs/hep-ph/0109064}{{\ttfamily arXiv:hep-ph/0109064
  [hep-ph]}}.

\bibitem{Arnold:2001ms}
P.~B. Arnold, G.~D. Moore, and L.~G. Yaffe, ``{Photon emission from quark gluon
  plasma: Complete leading order results},''
  \href{http://dx.doi.org/10.1088/1126-6708/2001/12/009}{{\em JHEP} {\bfseries
  0112} (2001) 009},
\href{http://arxiv.org/abs/hep-ph/0111107}{{\ttfamily arXiv:hep-ph/0111107
  [hep-ph]}}.

\bibitem{Arnold:2002ja}
P.~B. Arnold, G.~D. Moore, and L.~G. Yaffe, ``{Photon and gluon emission in
  relativistic plasmas},''
  \href{http://dx.doi.org/10.1088/1126-6708/2002/06/030}{{\em JHEP} {\bfseries
  0206} (2002) 030},
\href{http://arxiv.org/abs/hep-ph/0204343}{{\ttfamily arXiv:hep-ph/0204343
  [hep-ph]}}.

\bibitem{Besak:2010fb}
D.~Besak and D.~Bodeker, ``{Hard Thermal Loops for Soft or Collinear External
  Momenta},'' \href{http://dx.doi.org/10.1007/JHEP05(2010)007}{{\em JHEP}
  {\bfseries 1005} (2010) 007},
\href{http://arxiv.org/abs/1002.0022}{{\ttfamily arXiv:1002.0022 [hep-ph]}}.

\bibitem{Arnold:2002zm}
P.~B. Arnold, G.~D. Moore, and L.~G. Yaffe, ``{Effective kinetic theory for
  high temperature gauge theories},''
  \href{http://dx.doi.org/10.1088/1126-6708/2003/01/030}{{\em JHEP} {\bfseries
  0301} (2003) 030},
\href{http://arxiv.org/abs/hep-ph/0209353}{{\ttfamily arXiv:hep-ph/0209353
  [hep-ph]}}.

\bibitem{Micha:2004bv}
R.~Micha and I.~I. Tkachev, ``{Turbulent thermalization},''
  \href{http://dx.doi.org/10.1103/PhysRevD.70.043538}{{\em Phys. Rev.}
  {\bfseries D70} (2004) 043538},
\href{http://arxiv.org/abs/hep-ph/0403101}{{\ttfamily arXiv:hep-ph/0403101
  [hep-ph]}}.

\bibitem{Berges:2008wm}
J.~Berges, A.~Rothkopf, and J.~Schmidt, ``{Non-thermal fixed points: Effective
  weak-coupling for strongly correlated systems far from equilibrium},''
  \href{http://dx.doi.org/10.1103/PhysRevLett.101.041603}{{\em Phys. Rev.
  Lett.} {\bfseries 101} (2008) 041603},
\href{http://arxiv.org/abs/0803.0131}{{\ttfamily arXiv:0803.0131 [hep-ph]}}.

\bibitem{Kofman:1995fi}
L.~Kofman, A.~D. Linde, and A.~A. Starobinsky, ``{Nonthermal phase transitions
  after inflation},'' \href{http://dx.doi.org/10.1103/PhysRevLett.76.1011}{{\em
  Phys.Rev.Lett.} {\bfseries 76} (1996) 1011--1014},
\href{http://arxiv.org/abs/hep-th/9510119}{{\ttfamily arXiv:hep-th/9510119
  [hep-th]}}.

\bibitem{Khlebnikov:1998sz}
S.~Khlebnikov, L.~Kofman, A.~D. Linde, and I.~Tkachev, ``{First order
  nonthermal phase transition after preheating},''
  \href{http://dx.doi.org/10.1103/PhysRevLett.81.2012}{{\em Phys.Rev.Lett.}
  {\bfseries 81} (1998) 2012--2015},
\href{http://arxiv.org/abs/hep-ph/9804425}{{\ttfamily arXiv:hep-ph/9804425
  [hep-ph]}}.

\bibitem{Tkachev:1995md}
I.~Tkachev, ``{Phase transitions at preheating},''
  \href{http://dx.doi.org/10.1016/0370-2693(96)00297-3}{{\em Phys.Lett.}
  {\bfseries B376} (1996) 35--40},
\href{http://arxiv.org/abs/hep-th/9510146}{{\ttfamily arXiv:hep-th/9510146
  [hep-th]}}.

\bibitem{Felder:2000sf}
G.~N. Felder, L.~Kofman, A.~D. Linde, and I.~Tkachev, ``{Inflation after
  preheating},'' \href{http://dx.doi.org/10.1088/1126-6708/2000/08/010}{{\em
  JHEP} {\bfseries 0008} (2000) 010},
\href{http://arxiv.org/abs/hep-ph/0004024}{{\ttfamily arXiv:hep-ph/0004024
  [hep-ph]}}.

\bibitem{Yanagida:1979as}
T.~Yanagida, ``{HORIZONTAL SYMMETRY AND MASSES OF NEUTRINOS},''
{\em Conf.Proc.} {\bfseries C7902131} (1979) 95--99.

\bibitem{Yanagida:1980xy}
T.~Yanagida, ``{Horizontal Symmetry and Masses of Neutrinos},''
\href{http://dx.doi.org/10.1143/PTP.64.1103}{{\em Prog.Theor.Phys.} {\bfseries
  64} (1980) 1103}.

\bibitem{GellMann:1980vs}
M.~Gell-Mann, P.~Ramond, and R.~Slansky, ``{Complex Spinors and Unified
  Theories},'' {\em Conf.Proc.} {\bfseries C790927} (1979) 315--321,
\href{http://arxiv.org/abs/1306.4669}{{\ttfamily arXiv:1306.4669 [hep-th]}}.

\bibitem{Minkowski:1977sc}
P.~Minkowski, ``{$\mu \to e\gamma$ at a Rate of One Out of $10^{9}$ Muon
  Decays?},''
\href{http://dx.doi.org/10.1016/0370-2693(77)90435-X}{{\em Phys.Lett.}
  {\bfseries B67} (1977) 421--428}.

\bibitem{Vilenkin:1982ks}
A.~Vilenkin and A.~Everett, ``{Cosmic Strings and Domain Walls in Models with
  Goldstone and PseudoGoldstone Bosons},''
\href{http://dx.doi.org/10.1103/PhysRevLett.48.1867}{{\em Phys.Rev.Lett.}
  {\bfseries 48} (1982) 1867--1870}.

\bibitem{Kim:1979if}
J.~E. Kim, ``{Weak Interaction Singlet and Strong CP Invariance},''
\href{http://dx.doi.org/10.1103/PhysRevLett.43.103}{{\em Phys.Rev.Lett.}
  {\bfseries 43} (1979) 103}.

\bibitem{Shifman:1979if}
M.~A. Shifman, A.~Vainshtein, and V.~I. Zakharov, ``{Can Confinement Ensure
  Natural CP Invariance of Strong Interactions?},''
\href{http://dx.doi.org/10.1016/0550-3213(80)90209-6}{{\em Nucl.Phys.}
  {\bfseries B166} (1980) 493}.

\bibitem{Lazarides:1982tw}
G.~Lazarides and Q.~Shafi, ``{Axion Models with No Domain Wall Problem},''
\href{http://dx.doi.org/10.1016/0370-2693(82)90506-8}{{\em Phys.Lett.}
  {\bfseries B115} (1982) 21}.

\bibitem{Kawasaki:2015ofa}
M.~Kawasaki, M.~Yamada, and T.~T. Yanagida, ``{Observable dark radiation from a
  cosmologically safe QCD axion},''
  \href{http://dx.doi.org/10.1103/PhysRevD.91.125018}{{\em Phys.Rev.}
  {\bfseries D91} no.~12, (2015) 125018},
\href{http://arxiv.org/abs/1504.04126}{{\ttfamily arXiv:1504.04126 [hep-ph]}}.

\bibitem{Preskill:1982cy}
J.~Preskill, M.~B. Wise, and F.~Wilczek, ``{Cosmology of the Invisible
  Axion},''
\href{http://dx.doi.org/10.1016/0370-2693(83)90637-8}{{\em Phys.Lett.}
  {\bfseries B120} (1983) 127--132}.

\bibitem{Abbott:1982af}
L.~Abbott and P.~Sikivie, ``{A Cosmological Bound on the Invisible Axion},''
\href{http://dx.doi.org/10.1016/0370-2693(83)90638-X}{{\em Phys.Lett.}
  {\bfseries B120} (1983) 133--136}.

\bibitem{Dine:1982ah}
M.~Dine and W.~Fischler, ``{The Not So Harmless Axion},''
\href{http://dx.doi.org/10.1016/0370-2693(83)90639-1}{{\em Phys.Lett.}
  {\bfseries B120} (1983) 137--141}.

\bibitem{Fukugita:1986hr}
M.~Fukugita and T.~Yanagida, ``{Baryogenesis Without Grand Unification},''
\href{http://dx.doi.org/10.1016/0370-2693(86)91126-3}{{\em Phys.Lett.}
  {\bfseries B174} (1986) 45}.

\bibitem{Buchmuller:2005eh}
W.~Buchmuller, R.~Peccei, and T.~Yanagida, ``{Leptogenesis as the origin of
  matter},''
  \href{http://dx.doi.org/10.1146/annurev.nucl.55.090704.151558}{{\em
  Ann.Rev.Nucl.Part.Sci.} {\bfseries 55} (2005) 311--355},
\href{http://arxiv.org/abs/hep-ph/0502169}{{\ttfamily arXiv:hep-ph/0502169
  [hep-ph]}}.

\bibitem{Kuzmin:1985mm}
V.~Kuzmin, V.~Rubakov, and M.~Shaposhnikov, ``{On the Anomalous Electroweak
  Baryon Number Nonconservation in the Early Universe},''
\href{http://dx.doi.org/10.1016/0370-2693(85)91028-7}{{\em Phys.Lett.}
  {\bfseries B155} (1985) 36}.

\bibitem{Harigaya:2014tla}
K.~Harigaya, A.~Kamada, M.~Kawasaki, K.~Mukaida, and M.~Yamada, ``{Affleck-Dine
  Baryogenesis and Dark Matter Production after High-scale Inflation},''
  \href{http://dx.doi.org/10.1103/PhysRevD.90.043510}{{\em Phys.Rev.}
  {\bfseries D90} no.~4, (2014) 043510},
\href{http://arxiv.org/abs/1404.3138}{{\ttfamily arXiv:1404.3138 [hep-ph]}}.

\bibitem{Arnold:2008zu}
P.~B. Arnold and C.~Dogan, ``{QCD Splitting/Joining Functions at Finite
  Temperature in the Deep LPM Regime},''
  \href{http://dx.doi.org/10.1103/PhysRevD.78.065008}{{\em Phys. Rev.}
  {\bfseries D78} (2008) 065008},
\href{http://arxiv.org/abs/0804.3359}{{\ttfamily arXiv:0804.3359 [hep-ph]}}.

\end{thebibliography}\endgroup
  
\end{document}